\documentclass[aps,prl,reprint,showpacs,superscriptaddress,longbibliography]{revtex4-2}
\usepackage{graphicx}

\makeatletter
\AtBeginDocument{\let\LS@rot\@undefined}
\makeatother

\usepackage{amsmath,amssymb,amsfonts,braket}
\usepackage{comment}
\usepackage{color}
\usepackage[T1]{fontenc}
\usepackage{textcomp}
\usepackage{hyperref}
\hypersetup{
setpagesize=false,
colorlinks=true,
linkcolor=blue,
citecolor=red,
}
\usepackage{bm}
\usepackage{mathrsfs}
\usepackage{physics}
\usepackage{svg}
\usepackage{multirow}
\usepackage{pdfpages}
\usepackage{pgffor}

\def\supplementfilename{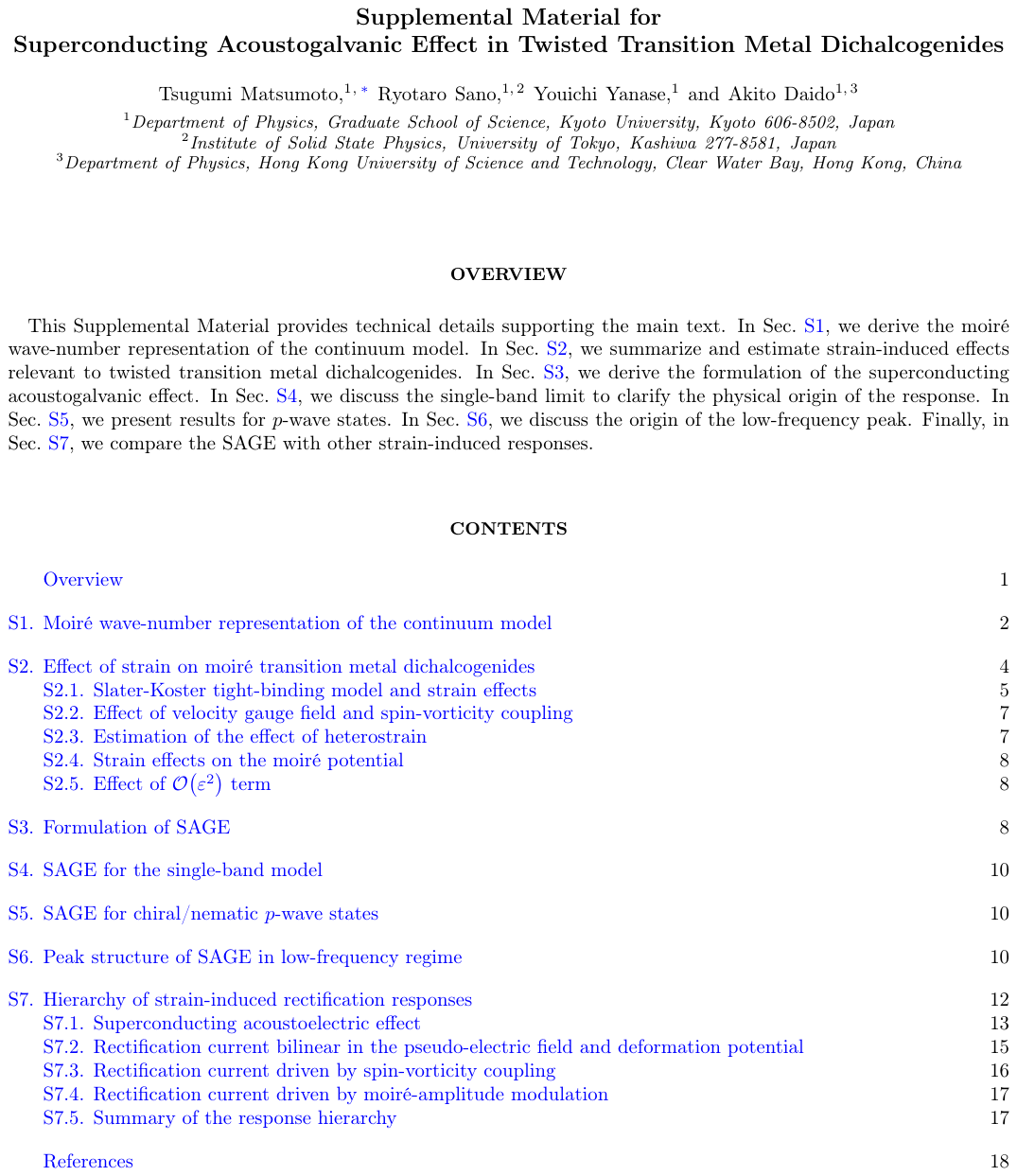}

\pdfximage{\supplementfilename}
\def\numbersupplementpages{\the\pdflastximagepages}

\newif\ifarXiv
\arXivtrue

\makeatletter
\let\MYcaption\@makecaption
\makeatother
\usepackage{subcaption}
\makeatletter
\let\@makecaption\MYcaption
\makeatother

\begin{document}

\preprint{APS/123-QED}

\title{Superconducting Acoustogalvanic Effect in 
Twisted Transition Metal Dichalcogenides
}

\author{Tsugumi Matsumoto}
\email{matsumoto.tsugumi.78w@st.kyoto-u.ac.jp}
\affiliation{Department of Physics, Graduate School of Science, Kyoto University, Kyoto 606-8502, Japan}

\author{Ryotaro Sano}
\affiliation{Department of Physics, Graduate School of Science, Kyoto University, Kyoto 606-8502, Japan}
\affiliation{Institute of Solid State Physics, University of Tokyo, Kashiwa 277-8581, Japan}

\author{Youichi Yanase}
\affiliation{Department of Physics, Graduate School of Science, Kyoto University, Kyoto 606-8502, Japan}

\author{Akito Daido}
\affiliation{Department of Physics, Graduate School of Science, Kyoto University, Kyoto 606-8502, Japan}
\affiliation{Department of Physics, Hong Kong University of Science and Technology, Clear Water Bay, Hong Kong, China}

\date{\today}

\begin{abstract}
Two-dimensional van der Waals superconductors are attracting much attention owing to their rich phase diagrams including possible unconventional superconductivity. However, they suffer from a lack of reliable methods for identifying their nontrivial  pairing symmetries and quantum geometry. In this study, we propose nonlinear responses driven by surface acoustic waves as a novel probe to access 
exotic Bogoliubov quasiparticles in such superconductors. Our approach is particularly suitable for addressing the superconducting gap structure as the gap energies in these systems typically lie within the frequency range of surface acoustic waves, and thus paves the way toward the experimental identification of exotic superconducting states especially in low-$T_c$ superconductors.
\end{abstract}
\maketitle

\textit{Introduction}.--- 
The recent discovery of twisted multilayer van der Waals (vdW) superconductors with the honeycomb structure, exemplified by magic angle twisted bilayer graphene~\cite{Cao2018-pn} and twisted WSe$_2$ (tWSe$_2$)~\cite{Xia2025-fp,Guo2025-kt}, has opened avenues for realizing various exotic superconducting states that have long been sought after.
Particularly in tWSe$_2$, chiral and nematic superconductivity---which spontaneously break time-reversal and rotational symmetries, respectively---are promising candidates for its observed superconducting state~\cite{Belanger2022-oa,Klebl2023-qr,Wu2023-oy,Schrade2024-cx,Guerci2024-zc}. Towards their experimental identification, it is essential
to clarify the wave-function properties of Cooper pairs and Bogoliubov quasiparticles, namely their symmetry and quantum geometry.
However, conventional probes such as optical responses and heat transport face difficulty in accessing these properties due to extremely low critical temperatures as well as the two-dimensional nature.
Therefore, developing advanced measurement methods more suitable for atomically thin vdW superconductors is highly desired.

Nonlinear responses have recently established their profound connections to the symmetry and quantum geometry inherent in normal-state Bloch electrons~\cite{Tokura2018-yp,Nagaosa2024-uk}. This concept has now been extended to superconducting states with particular emphasis on nonlinear optical responses~\cite{Wang2021-bq,Watanabe2022-pk,Watanabe2022-af,Tanaka2023-st,Tanaka2024-rt,Tanaka2025-dg,Kaplan2025-mz,Matsyshyn2026-ol}.
By activating the hidden quantum-geometric properties of Bogoliubov quasiparticles, nonlinear responses may provide a powerful tool for studying exotic superconducting states. However, the electromagnetic fields suffer from a fundamental limitation intrinsic to superconductors: the coupling between the vector potential and momentum enters with opposite signs in the particle and hole sectors. 
This fact sharply contrasts with the normal-state counterpart, and thus prevents nonlinear optical responses from directly capturing the quantum geometry of Bogoliubov quasiparticles in momentum space.

Surface acoustic waves (SAWs) are highly directional mechanical vibrations that propagate along the surface of elastic media. SAWs serve as a vibrational probe at the nanoscale, offering a compelling platform for simultaneously modulating materials via surfing carriers in 2D materials through piezoelectric and strain fields~\cite{Fal-ko1993-dz,Miseikis2012-zp,Bandhu2013-qq,Sonowal2020-bt,Mou2025-dm}. Specifically in honeycomb systems, a spatial modulation of the hopping energies due to strain mimics the role of artificial gauge fields~\cite{von-Oppen2009-at,Vaezi2013-uc,Cazalilla2014-eh,Kalameitsev2019-os,Sukhachov2020-ri,Sela2020-gl,Li2020-ch,Zhao2022-hr,Bhalla2022-su}, exerting a driving force even on charge-neutral quasiparticles~\cite{Uchoa2013-yi,Massarelli2017-sc,Nica2018-wy,Nayga2019-jx,Sano2024-gy,Yamazaki2023-av}. Therefore, the strain-induced gauge fields have the potential to replace the role of electromagnetic fields and temperature gradients in the study of atomically thin vdW materials, owing to their remarkable mechanical flexibility.
Furthermore, these strain gauge fields work at the two valley points in the opposite direction, thereby activating the valley degrees of freedom. 
This valley-contrasting nature offers a distinct advantage in overcoming the electromagnetic limitation in honeycomb superconductors. It enables direct access to quantum geometry, especially in systems where Cooper pairs are formed between electrons located at the opposite valley points.

\begin{figure}[tbp]
    \centering
\includegraphics[scale=0.25,keepaspectratio]{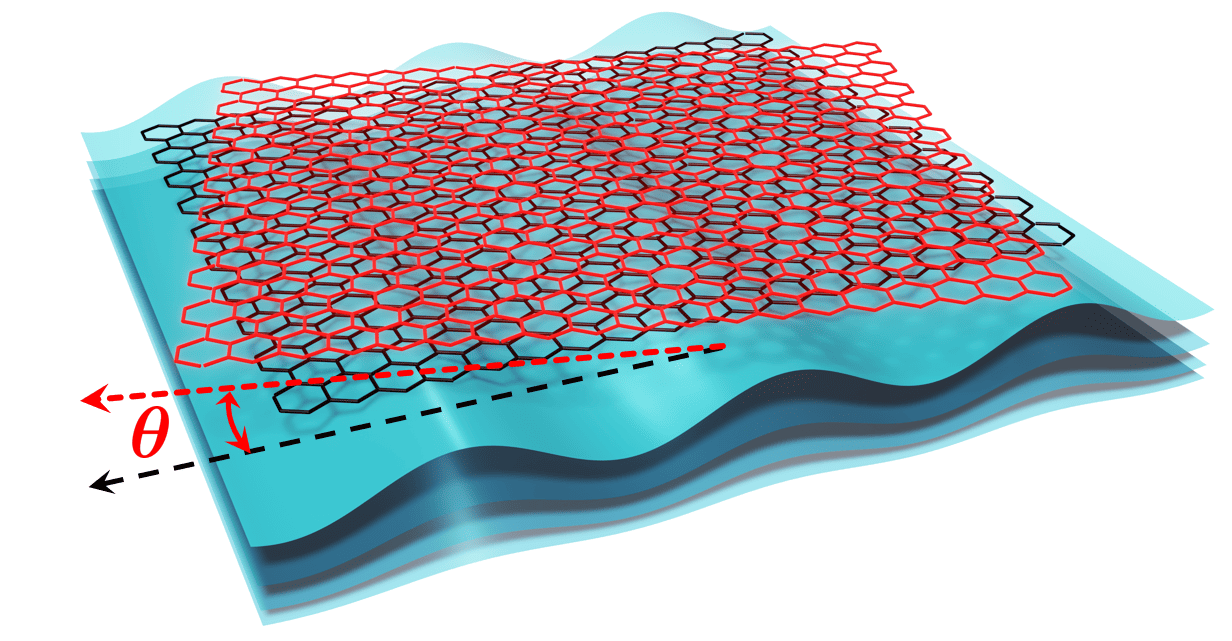}
    \caption{Schematics of the superconducting acoustogalvanic effect in twisted vdW materials with a twist angle $\theta$. A spatial modulation of the hopping energies due to strain mimics the role of artificial gauge fields for Bogoliubov quasiparticles in a honeycomb lattice.}
    \label{fig:tWSe2+SAW:image}
\end{figure}

In this Letter, we propose the \textit{superconducting acoustogalvanic effect} (SAGE) ---a nonlinear response driven by SAWs---as a novel probe of the pairing symmetry and quantum geometry of Bogoliubov quasiparticles in two-dimensional honeycomb superconductors including those formed with a twisted bilayer (Fig.~\ref{fig:tWSe2+SAW:image}).
The unique combination of mechanical tunability and valley-contrasting pseudo-gauge fields 
elevates SAWs as a promising tool for probing the symmetry and quantum-geometry aspects in atomically thin superconductors.
In particular, we focus on inter-valley pairing in tWSe$_2$, where the valley-contrasting nature of strain fields allows selective activation of quantum-geometric responses that are otherwise hidden in conventional electromagnetic probes.
We show that SAWs can induce a pronounced nonlinear response, tightly linked to the underlying quantum geometry of the superconducting state,
and further discuss its potential as a remarkable tool for investigating exotic superconducting states in vdW materials.

\textit{Model}.--- 
\begin{figure}[tbp]
    \centering
\includegraphics[scale=0.3,keepaspectratio]{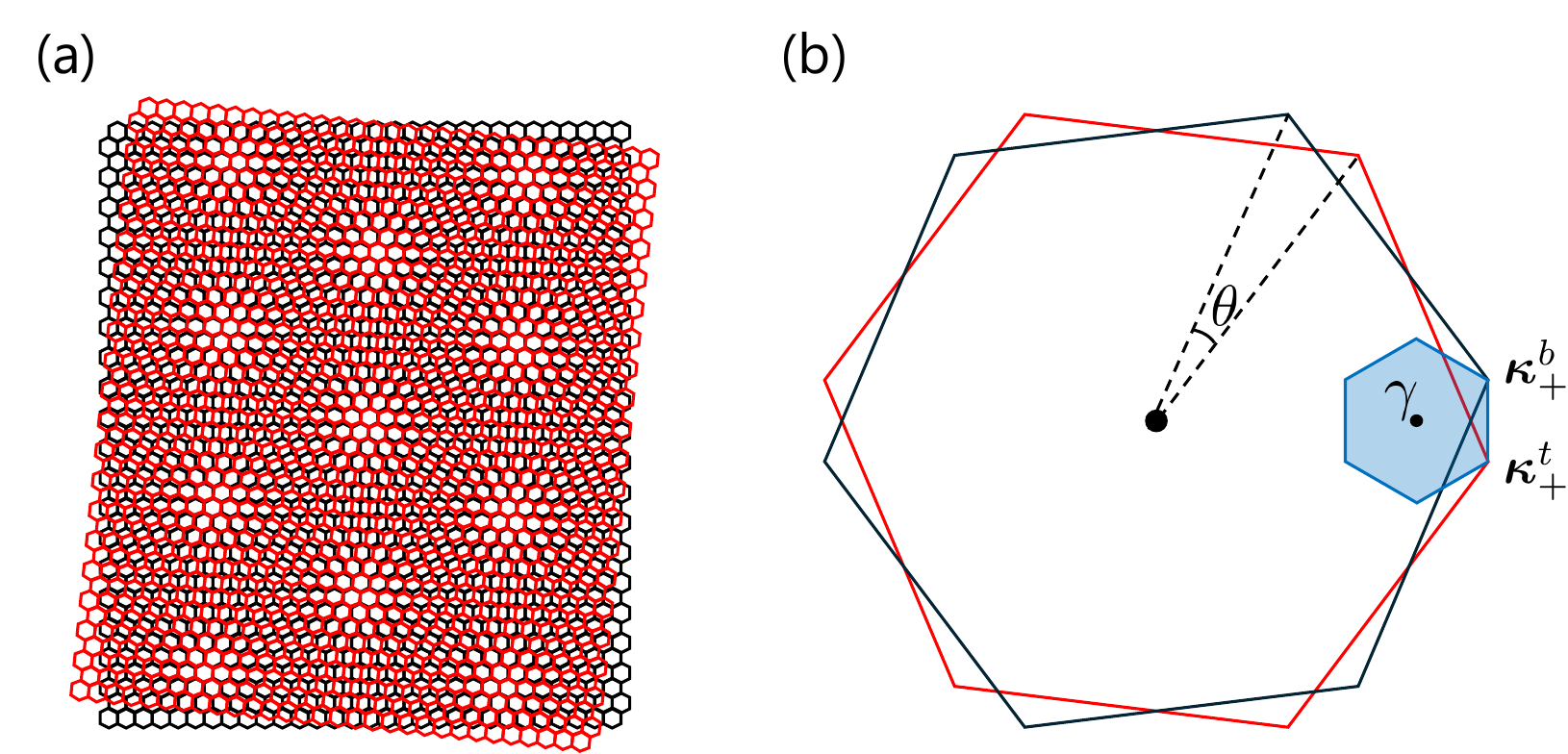}
    \caption{Schematics of (a) the moir\'e superlattice and (b) the corresponding reciprocal space with a twist angle $\theta$.
   The red and black sheets represent the top and bottom layers, respectively. The blue area is the folded moir\'e Brillouin zone.}
    \label{fig:moire_lattic+mBZ:image}
\end{figure}
To investigate SAWs responses in honeycomb superconductors, we consider an AA-stacked bilayer WSe$_2$ with a small twist angle $\theta$ (Fig.~\ref{fig:moire_lattic+mBZ:image}(a)).
We begin with the standard continuum model~\cite{Wu2019-ry} defined in the moir\'e Brillouin zone (MBZ) as depicted in Fig.~\ref{fig:moire_lattic+mBZ:image}(b),
\begin{equation}
    \hat{H}_{\tau} =
    \mqty(
        {H}_{\tau,b}(\hat{\bm{k}}) + \Delta_{b}(\hat{\bm{r}}) & T_{\tau}(\hat{\bm{r}}) \\
        T_{\tau}^{\dagger}(\hat{\bm{r}}) & {H}_{\tau,t}(\hat{\bm{k}}) + \Delta_{t}(\hat{\bm{r}})
    ),\label{Eq:continuum}
\end{equation}
where $\tau = \pm$ and $l = b/t = \pm$ denote the valley and layer degrees of freedom.
Owing to the spin-orbit coupling, the spin of the $\tau=\pm$ valley is locked to $\uparrow$ and $\downarrow$, respectively, and is kept implicit.
Here, ${H}_{\tau,l}(\hat{\bm{k}})$ is the Hamiltonian of each monolayer, and $\Delta_{l}(\hat{\bm{r}})$ and $T_\tau(\hat{\bm{r}})$ are intralayer and interlayer potentials, respectively.
We adopt the model parameters obtained from experimental configuration and density functional theory calculations~\cite{Devakul2021-hs,Guo2025-kt,Zhu2025-ba}. Further details of the model are provided in End Matter~\cite{Endmatter}.

To describe the superconducting state, we construct the unstrained Bogoliubov-de Gennes (BdG) Hamiltonian, 
\begin{align}
    \hat{H}_{\rm BdG} = \mqty(
    \hat{H}_+ & \Delta(\hat{\bm{k}}) \\
    \Delta^{\dagger}(\hat{\bm{k}}) & -\hat{H}_{-}^{T}
    ),\label{eq:BdG3}
\end{align}
whose eigenstates and eigenvalues are evaluated by keeping only the first-order interlayer coupling to calculate response functions~\cite{Supplemental}.
Various pairing symmetries have been proposed for tWSe$_2$, including chiral/nematic \( p \)- and \( d \)-wave states~\cite{Belanger2022-oa,Klebl2023-qr,Wu2023-oy,Schrade2024-cx,Guerci2024-zc} and a pair-density-wave (PDW) state~\cite{Wu2023-oy,Klebl2023-qr}.
In the main text, we consider both the chiral and nematic \( d \)-wave states as promising candidates of superconductivity in tWSe$_2$~\cite{Belanger2022-oa,Klebl2023-qr,Wu2023-oy}.
The corresponding order parameters are given by $\Delta(\bm{k})=\Delta(T)(k_x^2-k_y^2+2ik_xk_y)/k^2$ and $\Delta(T)(k_x^2-k_y^2)/k^2$, respectively, with $k=|\bm{k}|$.
The gap amplitude is assumed to follow a phenomenological temperature dependence
$\Delta(T) = \Delta_{0}\tanh\qty[1.74\sqrt{\frac{T_c}{T}-1}]$  with $\Delta_{0} = 1.76\,k_{\rm B}T_{\rm c}$, where $T_{\rm c}=0.426$\,K is the critical temperature  for the twist angle $5.0^\circ$~\cite{Guo2025-kt}.
The chiral/nematic \( p \)-wave state~\cite{Schrade2024-cx,Guerci2024-zc} is discussed in Sec.~S5 of the Supplemental Material and shown to exhibit qualitatively similar behavior to the \( d \)-wave states.
The case of PDW state is left as an intriguing future issue since it involves the intra-valley pairing and its SAW responses will probe information similar to the optical responses. This, in turn, may allow SAWs to distinguish the PDW state from nematic and chiral $p$- and $d$-wave states.

\textit{SAWs and pseudo-gauge fields}.---
To discuss superconducting SAWs responses, we consider SAWs propagating along the surface of the tWSe$_2$, which introduce a spatio-temporally periodic strain in the crystal lattice, as shown in Fig.~\ref{fig:tWSe2+SAW:image}. 
In particular, we focus on the Rayleigh-type SAWs, which can be excited under traction-free boundary conditions on piezoelectric substrates~\cite{Nie2023-cd}.
The Rayleigh-type SAWs propagating on the surface of piezoelectric substrates in the $xy$ plane are described by the displacement field
\begin{align}
    \bm{u}(\bm{r},t) = \Re\qty[\qty(u_{L}{\hat{Q}}+iu_{z}\hat{z})e^{i(\bm{Q}\cdot\bm{r}-\Omega t)}].
\end{align}
Here, $u_L$ and $u_z$ are the longitudinal and out-of-plane components, 
$\bm{Q} = Q\hat{\bm{Q}} = Q(\cos\varphi,\;\sin\varphi)$ is the in-plane propagating wave vector with $\varphi$ being an azimuthal angle,
$\Omega = c_t \xi Q$ is the frequency of SAWs applied to tWSe$_2$ and $\xi$ is a constant characterizing the SAWs dispersion~\cite{Landau1986-fa}.

Inhomogeneous strain modifies the electronic structure, thereby generating scalar and vector couplings in the low-energy effective Hamiltonian.
Here, we consider the leading valley-odd vector coupling, which takes the form of a pseudo-gauge field~\cite{Rostami2015-hh,von-Oppen2009-at,Vaezi2013-uc,Cazalilla2014-eh,Kalameitsev2019-os,Sukhachov2020-ri,Sela2020-gl,Li2020-ch,Zhao2022-hr,Bhalla2022-su,Uchoa2013-yi,Massarelli2017-sc,Nica2018-wy,Nayga2019-jx,Sano2024-gy,Hu2022-im}.
Assuming that the tWSe$_2$ on a piezoelectric substrate completely follows the displacement of the substrate, the Rayleigh-type SAWs-induced pseudo-gauge field reads
$\bm{A}_{s}(\bm{r},t) = -\frac{\hbar}{e} \frac{\beta}{2a_{0}}\qty(\varepsilon_{xx}-\varepsilon_{yy},\;-2\varepsilon_{xy})$,
where $\varepsilon_{ij} = (\partial_iu_j + \partial_ju_i + \partial_iu_z\partial_ju_z)/2$ represents the strain tensor.
This gives rise to the pseudo-electric field $\bm{E}_s(\bm{r},t)=-\partial_t\bm{A}_s(\bm{r},t)$ and the pseudo-magnetic field $\bm{B}_s(\bm{r},t)=\nabla\times\bm{A}_s(\bm{r},t)$, each of which oscillates along a certain direction
similarly to linearly-polarized light.
Note that the pseudo-gauge fields stemming from the out-of-plane displacement are proportional to $u_z^2$, which are less relevant under weak strain, and hence neglected in the following analysis.

The strain-induced pseudo-gauge field $\bm{A}_{s}$ is coupled to the valley degrees of freedom rather than charge degrees of freedom, working at the two valley points in the opposite direction: $H_{\tau,l}(\hat{\bm{k}})\to H_{\tau,l}(\hat{\bm{k}}+\frac{e}{\hbar}\tau\bm{A}_s(\hat{\bm{r}},t))$ in Eq.~\eqref{Eq:continuum}.
Its coupling to the BdG Hamiltonian is given by replacing $\hat{H}_{\tau}$ in Eq.~\eqref{eq:BdG3} accordingly.
To illustrate this, we consider moir\'e wave-number representation of $\hat{H}_{\rm BdG}$~\cite{Supplemental} by assuming a spatially uniform $\bm{A}_s$, for simplicity:
\begin{align}
H_{\rm BdG}(\bm{k})=\begin{pmatrix}
H_+(\bm{k}+\frac{e}{\hbar}\bm{A}_s)&\Delta(\bm{k})\\
\Delta^\dagger(\bm{k}) &-H_-^T(-\bm{k}-\frac{e}{\hbar}\bm{A}_s)\end{pmatrix}.\label{eq:HBdGk5}
\end{align}
This indicates that the pseudo-gauge field couples to the BdG Hamiltonian by $H_{\rm BdG}(\bm{k})\to H_{\rm BdG}(\bm{k}+\frac{e}{\hbar}\bm{A}_s)$, in the same way as the coupling of electromagnetic gauge fields to Bloch electrons, when the $\bm{k}$ dependence of $\Delta(\bm{k})$ is not significant~\cite{Supplemental}.
As a result, the quantum geometry of Bogoliubov quasiparticles is relevant to SAW-induced excitations.
This observation stands in sharp contrast to and illustrates the advantage of SAWs over optical responses, where the gauge field couples to $H_{\rm BdG}(\bm{k})$ with the opposite signs in the particle and hole sectors reflecting their negative and positive charges, and therefore cannot
directly capture the quantum geometry of Bogoliubov quasiparticles in momentum space.

To establish the microscopic basis of the strain coupling introduced above, we project a strained Slater-Koster model onto the relevant low-energy valence band of tWSe$_2$.
This analysis identifies the pseudo-gauge field as the leading valley-odd coupling and the deformation potential as the leading valley-even scalar coupling.
The derivation and classification of additional strain-induced couplings are summarized in the End Matter and detailed in Sec.~S2 and Table~S1 of the Supplemental Material~\cite{Supplemental}.
In the following, we focus on the pseudo-gauge-field contribution.

\textit{Superconducting acoustogalvanic effect}.---
The pseudo-electric field $\bm{E}_s$ is known to give rise to nonlinear responses in the normal state, referred to as the acoustogalvanic effect~\cite{Sukhachov2020-ri,Bhalla2022-su}. 
In this Letter, we propose its counterpart in the superconducting state, namely the SAGE as a probe of exotic superconductivity in two-dimensional materials. 
We define the SAGE as the rectification charge current driven by SAWs-induced pseudo-gauge fields:
\begin{align}
\langle{\hat{\mathcal{J}}^\alpha}\rangle=\chi^{\alpha;\beta\gamma}_{\bm{Q},-\bm{Q}}(0;\Omega,-\Omega)A_s^\beta(\bm{Q},\Omega)A_s^\gamma(-\bm{Q},-\Omega),\label{eq:chi_abc}
\end{align}
where $\langle{\hat{\mathcal{J}}^\alpha}\rangle$ represents the spatio-temporally uniform charge current and $\bm{A}_s(\bm{Q},\Omega)$ is a Fourier component of the pseudo-gauge field.
Note that the finite momentum transfer $\bm{Q}$ should be incorporated, although it is usually neglected for optical responses. This is because $Q$ is much larger than that for optical responses due to the difference in the speed of sound and light. 
Note also that only $(\beta,\gamma)$-symmetric components contribute among various components of $\chi^{\alpha;\beta\gamma}_{\bm{Q},-\bm{Q}}$, since $\bm{A}_s$ is linearly polarized.
In the following, we concentrate on the propagation directions $\varphi=0$ and $2\pi/3$ where the pseudo-magnetic field vanishes, and rewrite Eq.~\eqref{eq:chi_abc} in the form of conductivity,
\begin{align}
\langle{\hat{\mathcal{J}}^\alpha}\rangle=\sigma^{\alpha;\beta\gamma}_{\bm{Q},-\bm{Q}}(0;\Omega,-\Omega)E_s^\beta(\bm{Q},\Omega)E_s^\gamma(-\bm{Q},-\Omega),
\end{align}
to highlight the similarity to the optical responses.

Similarly to the nonlinear optical responses~\cite{Ahn2020-ee,Watanabe2021-cf},
the acoustogalvanic conductivity $\sigma^{\alpha;\beta\gamma}_{\bm{Q},-\bm{Q}}(0;\Omega,-\Omega)$ is divided into several contributions with different physical origins.
In this Letter, we focus on the contribution that corresponds to the injection current, which is expected to be dominant in the clean-limit superconductors and therefore relevant to unconventional superconductivity.
By generalizing the derivation of nonlinear optical responses~\cite{Watanabe2022-pk}, this contribution, $\sigma_{\rm inj}^{\alpha;\beta\gamma}$, is given by
\begin{align}
    \sigma^{\alpha;\beta\gamma}_{\rm inj}
    &= \frac{e^2\pi}{2\hbar\Gamma}\int\frac{\dd[2]{\vb*{k}}}{(2\pi)^2}\sum_{a,b}(\mathcal{J}^\alpha_{aa}-\mathcal{J}^\alpha_{bb})G^{\beta\gamma}_{ab} F_{ab}(\Omega).
    \label{eq:inj}
\end{align}
Here, the indices $a$ and $b$ specify the eigenstates of $\hat{H}_{\rm BdG}$ with the wave numbers $\bm{k}+\bm{Q}/2$ and $\bm{k}-\bm{Q}/2$, respectively, while $E_{ab} = E_{a} - E_{b}$, $F_{ab}(\Omega) = [f(E_{b})-f(E_{a})]\delta(\hbar\Omega-E_{ab})$, and $G^{\beta\gamma}_{ab}=(\hbar/e)^2(\mathcal{V}_{ab}^{\beta}\mathcal{V}^{\gamma}_{ba}+\mathcal{V}_{ab}^{\gamma}\mathcal{V}^{\beta}_{ba})/{2E_{ab}^2}$, with the eigenvalue $E_a$ and Fermi-Dirac distribution function $f(E)=(e^{E/k_{\rm B}T}+1)^{-1}$.
The $\bm{k}$ integral runs over the MBZ.
The scattering rate \(\Gamma\) was introduced by replacing the adiabaticity parameter, following the formulation of the injection-current optical responses~\cite{Watanabe2022-pk,Supplemental}.

The matrix elements such as $\mathcal{J}_{aa}^\alpha$ and $\mathcal{V}_{ab}^\beta$ represent those for the charge current operator $\hat{\mathcal{J}}^\alpha$ and its counterpart for the pseudo-gauge field, $\hat{\mathcal{V}}^\beta=\lim_{\bm{A}_s\to0}(-\partial_{A_s^\beta}\hat{H}_{\rm BdG})$, namely the valley-current operator.
The precise expression of $\mathcal{V}^\beta_{ab}$ with momentum transfer $\bm{Q}$ is available in Sec.~S3 of the Supplemental Material~\cite{Supplemental}.
We can replace $\partial_{A_s^\beta}$ with $\partial_{k_\beta}$ when 
$\partial_{k_\beta}\Delta(\bm{k})$ makes negligible contribution to $G_{ab}^{\beta\gamma}$, and then $G_{ab}^{\beta\gamma}$ for $\bm{Q}\to0$ coincides with the band-resolved quantum metric of Bogoliubov quasiparticles.
This replacement is confirmed to be a good approximation for our model of tWSe$_2$, while we use the rigorous expression for numerical calculations. 
More generally, replacing the pseudo-gauge-field vertex by another strain-coupling vertex defines a corresponding coupling-dependent quantum-geometric factor, as discussed for the deformation potential in the End Matter~\cite{Endmatter}.

We can estimate the magnitude of $\sigma_{\rm inj}^{\alpha;\beta\gamma}$ by considering a parabolic band and a constant order parameter.
We then obtain $\sigma^{\alpha;\beta\gamma}_{\rm inj}=O(\xi_0^2Q/\Gamma)$ with the coherence length $\xi_0$~\cite{Supplemental}.
Thus, a finite SAGE response is expected regardless of the pairing symmetry, and serves as a versatile probe of superconducting states in vdW superconductors.
In particular, neither inversion nor time-reversal-symmetry breaking is required due to a finite momentum transfer $\bm{Q}\neq0$, in contrast to the $(\beta,\gamma)$-symmetric optical injection current~\cite{Ahn2020-ee,Watanabe2021-cf,Watanabe2022-pk}.

\textit{Applications to tWSe$_2$}.--- 
We are now ready to investigate the SAGE responses in tWSe$_2$. In particular, we discuss the frequency dependence of the acoustogalvanic conductivity Eq.~\eqref{eq:inj}.
Before evaluating these quantities, we analyze the joint density of states (JDOS), $J(\Omega;\hat{Q}) = \sum_{a,b}\int\frac{\dd[2]{\vb*{k}}}{(2\pi)^2}F_{ab}(\Omega)$, which quantifies the number of available resonance transitions at a frequency $\Omega$ between the states $a$ and $b$ with the wave numbers $\bm{k}+\bm{Q}/2$ and $\bm{k}-\bm{Q}/2$, respectively.
JDOS serves as a qualitative guide to the behavior of the response across different frequencies and temperatures.

\begin{figure*}[tbp]
    \centering
    \includegraphics[width=\linewidth,keepaspectratio]{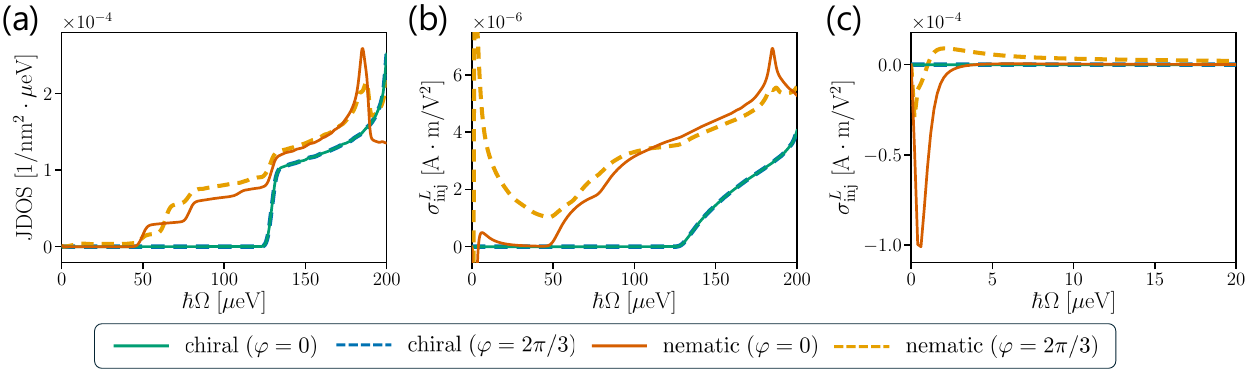}
    \caption{Frequency dependence of JDOS and acoustogalvanic conductivity $\sigma^L_{\mathrm{inj}}$. Green (orange) curves denote chiral (nematic) states.
    In (a)--(c), solid (dashed) lines represent the azimuthal angle $\varphi=0$ ($2\pi/3$); results for chiral states coincide due to symmetry.
    Panel (c) shows a magnified view of panel (b) in the low-frequency range $0 \leq \Omega/2\pi \leq 5\,\mathrm{GHz}$, corresponding to $0 \leq \hbar\Omega \leq 20\,\mathrm{\mu eV}$.
    Parameters are fixed at $T=50\,\mathrm{mK}$. Chiral states show a sharp increase near $\hbar\Omega = 2\Delta(T)$, whereas nematic states respond even below $2\Delta(T)$ due to low-energy quasiparticles.}
    \label{fig:plot_d}
\end{figure*}
\begin{figure}[tbp]
    \centering
    \includegraphics[width=\linewidth,keepaspectratio]{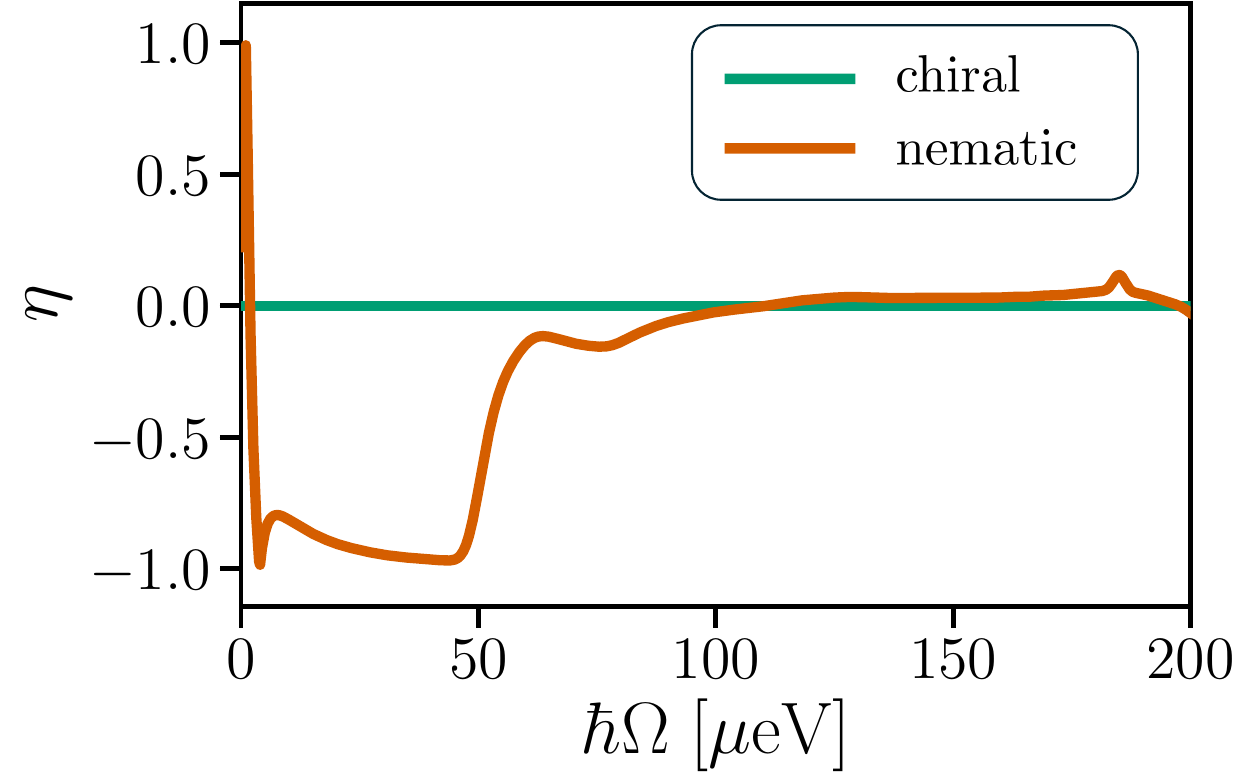}
    \caption{Frequency dependence of the nematicity $\eta$.
    Green and orange curves denote the chiral and nematic states, respectively.
    The temperature is fixed at $T=50\,\mathrm{mK}$.
    The nematicity vanishes in the chiral state because of $C_3$ symmetry, whereas it is finite in the nematic state owing to spontaneous rotational symmetry breaking.}
    \label{fig:plot_d_nematicity}
\end{figure}

The numerical results of JDOS for chiral and nematic superconducting states are shown in Fig.~\ref{fig:plot_d}(a).
Throughout, the SAW wave number $Q$ and frequency $\Omega$ are related by the dispersion relation $\Omega = c_t \xi Q$.
To understand the excitations for $\varphi=0$ and $2\pi/3$, we fixed temperature $T = 50\;\mathrm{mK}$, which lies well below the critical temperature $T_c=0.426\;\mathrm{K}$ and thus $\Delta(T)\simeq \Delta_0$.
Since the superconducting gap is estimated to be $\Delta_0 \simeq 64\; \mu\mathrm{eV}$ in tWSe$_2$, the resonance at $\hbar\Omega = 2\Delta_0$ is manifested as the typical frequency where JDOS begins to have sizable values.
In the chiral state, where a fully-gapped state is realized, JDOS almost vanishes below the resonance and exhibits a sharp increase at $\hbar\Omega \simeq 2\Delta_0$.
On the other hand, in the nematic state, a finite JDOS persists even below the resonance frequency due to its nodal and anisotropic gap structure.
In this case, JDOS depends on the propagation direction $\varphi$, as a result of the nematicity.

Based on the insight of JDOS, we discuss the SAGE in tWSe$_2$.
For SAWs directed along $\varphi=0$ and $2\pi/3$, $\bm{Q}$ and $\bm{E}_s$ are parallel, and we focus on the longitudinal conductivity $\sigma^L_{\rm inj}\equiv \hat{Q}_\alpha\sigma^{\alpha;\beta\gamma}_{\rm inj}\hat{Q}_\beta\hat{Q}_\gamma$, for the chiral and nematic states.
This corresponds to $\sigma^{x;xx}_{\rm inj}$ for $\varphi=0$, for instance.
Figures~\ref{fig:plot_d}(b) and (c) show the frequency dependence of the acoustogalvanic conductivity $\sigma^{L}_{\rm inj}$ for the chiral and nematic states at the same parameters as in Fig.~\ref{fig:plot_d}(a).
Here, Fig.~\ref{fig:plot_d}(c) is a magnified view of Fig.~\ref{fig:plot_d}(b).
As shown in Figs.~\ref{fig:plot_d}(b) and (c), the response of the chiral state is suppressed in the low-frequency regime due to the full-gap structure.
As the frequency approaches the threshold $\hbar\Omega=2\Delta_0$ for the quasiparticle excitations, $\sigma^{L}_{\rm inj}$ first shows a sharp increase, as in JDOS.
The nematic state exhibits finite responses even at frequencies smaller than $2\Delta_0$ owing to the presence of low-energy quasiparticles.
Interestingly, a significant low-frequency enhancement is obtained  for both $\varphi=0$ and $2\pi/3$.
As shown in Sec.~S6 of the Supplemental Material, this structure remains even with $\Delta(T)=0$ (i.e., in the normal state)~\cite{Supplemental}, indicating the importance of the resonances near the nodal directions, where the quasiparticles behave similarly to the normal state.
More precisely, since the JDOS shows no low-frequency divergence, the enhancement is attributed instead to the divergence of the quantum metric with finite momentum transfer $G_{ab}^{\beta\gamma}$ in Eq.~\eqref{eq:inj}~\cite{Supplemental}.
The results illustrate that SAGE conductivity can sensitively probe the low-energy quasiparticles of nodal superconducting states.

Furthermore, we investigate the nematicity of the acoustogalvanic conductivity, which is defined by $\eta \equiv\frac{\abs{\sigma^L_{\rm inj}(\varphi = 0)}-\abs{\sigma^L_{\rm inj}(\varphi = 2\pi/3)}}{\abs{\sigma^L_{\rm inj}(\varphi = 0)}+\abs{\sigma^L_{\rm inj}(\varphi = 2\pi/3)}}$.
In the normal and chiral states, the system possesses $C_3$ rotational symmetry, prohibiting a finite nematicity.
In contrast, the spontaneous rotational symmetry breaking in the nematic state allows a finite $\eta$, as clearly indicated in Fig.~\ref{fig:plot_d_nematicity}: The frequency dependence demonstrates that only the nematic state exhibits finite nematicity.
Therefore, the emergent nematicity below the critical temperature serves as a smoking gun for identifying the pairing symmetry, complementing the information obtained from frequency dependence of acoustogalvanic conductivity.
Thus, our formulation of SAGE opens a new avenue for identifying unconventional superconducting states in vdW materials and beyond.

Here, we discuss the temperature dependence of SAGE.
Numerical results and further details are presented in the End Matter~\cite{Endmatter}.
As in the frequency dependence, the temperature dependence of SAGE is governed primarily by the resonance condition $\hbar\Omega \simeq 2\Delta(T)$, as suggested by the JDOS analysis.
At a fixed SAW frequency, cooling increases the superconducting gap and therefore changes the available resonant quasiparticle transitions.
Consequently, in the fully gapped chiral state, the SAGE response is suppressed at low temperatures once $2\Delta(T)$ exceeds $\hbar\Omega$, whereas the nodal nematic state retains a finite response owing to low-energy quasiparticles.
In addition, only the nematic state exhibits finite nematicity, which serves as a smoking-gun signature of spontaneous rotational-symmetry breaking.
Thus, temperature-dependent measurements of SAGE provide a new probe of unconventional superconductivity in two-dimensional materials.

\textit{Discussion}.--- 
Finally, we discuss the experimental feasibility of the SAGE.
Under realistic experimental conditions, the amplitude of charge current is estimated to be of the order of $1\;{\rm nA/nm}$ with the pseudo-electric field $\abs{\vb*{E}_{s}}\simeq1.5\times10^{3}\;\mathrm{V/m}$ induced by the Rayleigh-type SAWs, as well as a phenomenological  scattering rate $\Gamma= \Delta_0/10$~\cite{Supplemental}.
Here, we have used the parameters~\cite{Fang2018-wv,Bi2019-aa,Maznev2021-ou}: 
$u_\mathrm{L}\simeq100\,\mathrm{pm}$, $c_t\simeq4000\,\mathrm{m/s}$, $\xi\simeq0.95$, $\beta\sim2.3$, and $\Omega/2\pi= 25\,\mathrm{GHz}$.
The corresponding wavelength $\lambda_\mathrm{SAW}\simeq160\;\mathrm{nm}$ is comparable to the coherence length of tWSe$_2$ ($\sim57\;\mathrm{nm}$~\cite{Guo2025-kt}), highlighting the significance of the momentum transfer \(\bm{Q}\) in SAGE.
Detailed experimental configuration for suppressing piezoelectric-field-induced backgrounds and detecting strain-induced dc responses are discussed in End Matter~\cite{Endmatter}.

In addition to the pseudo-gauge-field contribution, the valley-even deformation potential can generate a rectified dc current through the superconducting acoustoelectric effect (SAEE).
Depending on the screening environment and device geometry, the SAEE may become comparable to or even dominate over the SAGE.
Nevertheless, the low-frequency enhancement associated with nodal quasiparticles and the emergence of nematicity below $T_{\rm c}$ remain useful diagnostics of the superconducting gap structure, although identifying the response specifically as the SAGE requires control over the relevant strain-coupling channels.
Possible strategies for separating the two contributions, including gate-distance dependence and the use of transverse acoustic modes, are discussed in the End Matter and Sec.~S7 of the Supplemental Material~\cite{Supplemental, Endmatter}.

Most candidates of exotic superconductors exhibit low critical temperatures, typically ranging from 0.1 to 1\,K,
with superconducting gaps estimated to be on the order of $10^{1-2}\,\mu$eV.
This energy scale lies outside the accessible range of the conventional optical measurement. 
In contrast, SAWs operate in the MHz to GHz range, serving our SAGE as a powerful and suitable tool for probing such low-energy excitations and gap structures.
Momentum transfer \(\bm{Q}\) ---ignored in conventional optical responses--- is unique to SAWs.
This distinction shows that optical response and SAGE probe fundamentally different physics, thereby encoding the value of SAGE as a new experimental probe.

\begin{acknowledgments}
The authors are grateful to R. Hisatomi, M. Tanaka, Y. Niimi, K. Shinada, H. Tanaka, S. Asano, and Y. Hirobe for fruitful discussions.
R.~S. thanks Y. Ominato and M. Matsuo for their introduction to the strain gauge fields and SAWs. 
A.~D. thanks Z.-T. Sun for helpful comments on the manuscript.
This work was supported by JST SPRING (Grant Number JPMJSP2110) and JSPS KAKENHI (Grants No.JP22KJ1937, No.JP22H01181, No.JP22H04933, No.JP23K17353, No.JP23K22452, No.JP23KK0248, No.JP24H00007, No.JP24H01662, No.JP24K21530, No.JP25H01249, No.JP26H02016 and No.JP26KJ1377).
\end{acknowledgments}

\bibliography{ref}

\begin{thebibliography}{67}%
\makeatletter
\providecommand \@ifxundefined [1]{%
 \@ifx{#1\undefined}
}%
\providecommand \@ifnum [1]{%
 \ifnum #1\expandafter \@firstoftwo
 \else \expandafter \@secondoftwo
 \fi
}%
\providecommand \@ifx [1]{%
 \ifx #1\expandafter \@firstoftwo
 \else \expandafter \@secondoftwo
 \fi
}%
\providecommand \natexlab [1]{#1}%
\providecommand \enquote  [1]{``#1''}%
\providecommand \bibnamefont  [1]{#1}%
\providecommand \bibfnamefont [1]{#1}%
\providecommand \citenamefont [1]{#1}%
\providecommand \href@noop [0]{\@secondoftwo}%
\providecommand \href [0]{\begingroup \@sanitize@url \@href}%
\providecommand \@href[1]{\@@startlink{#1}\@@href}%
\providecommand \@@href[1]{\endgroup#1\@@endlink}%
\providecommand \@sanitize@url [0]{\catcode `\\12\catcode `\$12\catcode `\&12\catcode `\#12\catcode `\^12\catcode `\_12\catcode `\%12\relax}%
\providecommand \@@startlink[1]{}%
\providecommand \@@endlink[0]{}%
\providecommand \url  [0]{\begingroup\@sanitize@url \@url }%
\providecommand \@url [1]{\endgroup\@href {#1}{\urlprefix }}%
\providecommand \urlprefix  [0]{URL }%
\providecommand \Eprint [0]{\href }%
\providecommand \doibase [0]{https://doi.org/}%
\providecommand \selectlanguage [0]{\@gobble}%
\providecommand \bibinfo  [0]{\@secondoftwo}%
\providecommand \bibfield  [0]{\@secondoftwo}%
\providecommand \translation [1]{[#1]}%
\providecommand \BibitemOpen [0]{}%
\providecommand \bibitemStop [0]{}%
\providecommand \bibitemNoStop [0]{.\EOS\space}%
\providecommand \EOS [0]{\spacefactor3000\relax}%
\providecommand \BibitemShut  [1]{\csname bibitem#1\endcsname}%
\let\auto@bib@innerbib\@empty
\bibitem [{\citenamefont {Cao}\ \emph {et~al.}(2018)\citenamefont {Cao}, \citenamefont {Fatemi}, \citenamefont {Fang}, \citenamefont {Watanabe}, \citenamefont {Taniguchi}, \citenamefont {Kaxiras},\ and\ \citenamefont {Jarillo-Herrero}}]{Cao2018-pn}%
  \BibitemOpen
  \bibfield  {author} {\bibinfo {author} {\bibfnamefont {Y.}~\bibnamefont {Cao}}, \bibinfo {author} {\bibfnamefont {V.}~\bibnamefont {Fatemi}}, \bibinfo {author} {\bibfnamefont {S.}~\bibnamefont {Fang}}, \bibinfo {author} {\bibfnamefont {K.}~\bibnamefont {Watanabe}}, \bibinfo {author} {\bibfnamefont {T.}~\bibnamefont {Taniguchi}}, \bibinfo {author} {\bibfnamefont {E.}~\bibnamefont {Kaxiras}},\ and\ \bibinfo {author} {\bibfnamefont {P.}~\bibnamefont {Jarillo-Herrero}},\ }\bibfield  {title} {\bibinfo {title} {Unconventional superconductivity in magic-angle graphene superlattices},\ }\href {https://doi.org/10.1038/nature26160} {\bibfield  {journal} {\bibinfo  {journal} {Nature}\ }\textbf {\bibinfo {volume} {556}},\ \bibinfo {pages} {43} (\bibinfo {year} {2018})}\BibitemShut {NoStop}%
\bibitem [{\citenamefont {Xia}\ \emph {et~al.}(2025)\citenamefont {Xia}, \citenamefont {Han}, \citenamefont {Watanabe}, \citenamefont {Taniguchi}, \citenamefont {Shan},\ and\ \citenamefont {Mak}}]{Xia2025-fp}%
  \BibitemOpen
  \bibfield  {author} {\bibinfo {author} {\bibfnamefont {Y.}~\bibnamefont {Xia}}, \bibinfo {author} {\bibfnamefont {Z.}~\bibnamefont {Han}}, \bibinfo {author} {\bibfnamefont {K.}~\bibnamefont {Watanabe}}, \bibinfo {author} {\bibfnamefont {T.}~\bibnamefont {Taniguchi}}, \bibinfo {author} {\bibfnamefont {J.}~\bibnamefont {Shan}},\ and\ \bibinfo {author} {\bibfnamefont {K.~F.}\ \bibnamefont {Mak}},\ }\bibfield  {title} {\bibinfo {title} {Superconductivity in twisted bilayer {WSe$_2$}},\ }\href {https://doi.org/10.1038/s41586-024-08116-2} {\bibfield  {journal} {\bibinfo  {journal} {Nature}\ }\textbf {\bibinfo {volume} {637}},\ \bibinfo {pages} {833} (\bibinfo {year} {2025})}\BibitemShut {NoStop}%
\bibitem [{\citenamefont {Guo}\ \emph {et~al.}(2025)\citenamefont {Guo}, \citenamefont {Pack}, \citenamefont {Swann}, \citenamefont {Holtzman}, \citenamefont {Cothrine}, \citenamefont {Watanabe}, \citenamefont {Taniguchi}, \citenamefont {Mandrus}, \citenamefont {Barmak}, \citenamefont {Hone}, \citenamefont {Millis}, \citenamefont {Pasupathy},\ and\ \citenamefont {Dean}}]{Guo2025-kt}%
  \BibitemOpen
  \bibfield  {author} {\bibinfo {author} {\bibfnamefont {Y.}~\bibnamefont {Guo}}, \bibinfo {author} {\bibfnamefont {J.}~\bibnamefont {Pack}}, \bibinfo {author} {\bibfnamefont {J.}~\bibnamefont {Swann}}, \bibinfo {author} {\bibfnamefont {L.}~\bibnamefont {Holtzman}}, \bibinfo {author} {\bibfnamefont {M.}~\bibnamefont {Cothrine}}, \bibinfo {author} {\bibfnamefont {K.}~\bibnamefont {Watanabe}}, \bibinfo {author} {\bibfnamefont {T.}~\bibnamefont {Taniguchi}}, \bibinfo {author} {\bibfnamefont {D.~G.}\ \bibnamefont {Mandrus}}, \bibinfo {author} {\bibfnamefont {K.}~\bibnamefont {Barmak}}, \bibinfo {author} {\bibfnamefont {J.}~\bibnamefont {Hone}}, \bibinfo {author} {\bibfnamefont {A.~J.}\ \bibnamefont {Millis}}, \bibinfo {author} {\bibfnamefont {A.}~\bibnamefont {Pasupathy}},\ and\ \bibinfo {author} {\bibfnamefont {C.~R.}\ \bibnamefont {Dean}},\ }\bibfield  {title} {\bibinfo {title} {Superconductivity in 5.0° twisted bilayer {WSe$_2$}},\ }\href {https://doi.org/10.1038/s41586-024-08381-1} {\bibfield  {journal}
  {\bibinfo  {journal} {Nature}\ }\textbf {\bibinfo {volume} {637}},\ \bibinfo {pages} {839} (\bibinfo {year} {2025})}\BibitemShut {NoStop}%
\bibitem [{\citenamefont {Bélanger}\ \emph {et~al.}(2022)\citenamefont {Bélanger}, \citenamefont {Fournier},\ and\ \citenamefont {Sénéchal}}]{Belanger2022-oa}%
  \BibitemOpen
  \bibfield  {author} {\bibinfo {author} {\bibfnamefont {M.}~\bibnamefont {Bélanger}}, \bibinfo {author} {\bibfnamefont {J.}~\bibnamefont {Fournier}},\ and\ \bibinfo {author} {\bibfnamefont {D.}~\bibnamefont {Sénéchal}},\ }\bibfield  {title} {\bibinfo {title} {Superconductivity in the twisted bilayer transition metal dichalcogenide {WSe}$_2$ : A quantum cluster study},\ }\href {https://doi.org/10.1103/physrevb.106.235135} {\bibfield  {journal} {\bibinfo  {journal} {Phys. Rev. B.}\ }\textbf {\bibinfo {volume} {106}},\ \bibinfo {pages} {235135} (\bibinfo {year} {2022})}\BibitemShut {NoStop}%
\bibitem [{\citenamefont {Klebl}\ \emph {et~al.}(2023)\citenamefont {Klebl}, \citenamefont {Fischer}, \citenamefont {Classen}, \citenamefont {Scherer},\ and\ \citenamefont {Kennes}}]{Klebl2023-qr}%
  \BibitemOpen
  \bibfield  {author} {\bibinfo {author} {\bibfnamefont {L.}~\bibnamefont {Klebl}}, \bibinfo {author} {\bibfnamefont {A.}~\bibnamefont {Fischer}}, \bibinfo {author} {\bibfnamefont {L.}~\bibnamefont {Classen}}, \bibinfo {author} {\bibfnamefont {M.~M.}\ \bibnamefont {Scherer}},\ and\ \bibinfo {author} {\bibfnamefont {D.~M.}\ \bibnamefont {Kennes}},\ }\bibfield  {title} {\bibinfo {title} {Competition of density waves and superconductivity in twisted tungsten diselenide},\ }\href {https://doi.org/10.1103/physrevresearch.5.l012034} {\bibfield  {journal} {\bibinfo  {journal} {Phys. Rev. Res.}\ }\textbf {\bibinfo {volume} {5}},\ \bibinfo {pages} {L012034} (\bibinfo {year} {2023})}\BibitemShut {NoStop}%
\bibitem [{\citenamefont {Wu}\ \emph {et~al.}(2023)\citenamefont {Wu}, \citenamefont {Wu},\ and\ \citenamefont {Yao}}]{Wu2023-oy}%
  \BibitemOpen
  \bibfield  {author} {\bibinfo {author} {\bibfnamefont {Y.-M.}\ \bibnamefont {Wu}}, \bibinfo {author} {\bibfnamefont {Z.}~\bibnamefont {Wu}},\ and\ \bibinfo {author} {\bibfnamefont {H.}~\bibnamefont {Yao}},\ }\bibfield  {title} {\bibinfo {title} {{Pair-Density-Wave and Chiral Superconductivity in Twisted Bilayer Transition Metal Dichalcogenides}},\ }\href {https://doi.org/10.1103/physrevlett.130.126001} {\bibfield  {journal} {\bibinfo  {journal} {Phys. Rev. Lett.}\ }\textbf {\bibinfo {volume} {130}},\ \bibinfo {pages} {126001} (\bibinfo {year} {2023})}\BibitemShut {NoStop}%
\bibitem [{\citenamefont {Schrade}\ and\ \citenamefont {Fu}(2024)}]{Schrade2024-cx}%
  \BibitemOpen
  \bibfield  {author} {\bibinfo {author} {\bibfnamefont {C.}~\bibnamefont {Schrade}}\ and\ \bibinfo {author} {\bibfnamefont {L.}~\bibnamefont {Fu}},\ }\bibfield  {title} {\bibinfo {title} {Nematic, chiral, and topological superconductivity in twisted transition metal dichalcogenides},\ }\href {https://doi.org/10.1103/physrevb.110.035143} {\bibfield  {journal} {\bibinfo  {journal} {Phys. Rev. B.}\ }\textbf {\bibinfo {volume} {110}},\ \bibinfo {pages} {035143} (\bibinfo {year} {2024})}\BibitemShut {NoStop}%
\bibitem [{\citenamefont {Guerci}\ \emph {et~al.}(2024)\citenamefont {Guerci}, \citenamefont {Kaplan}, \citenamefont {Ingham}, \citenamefont {Pixley},\ and\ \citenamefont {Millis}}]{Guerci2024-zc}%
  \BibitemOpen
  \bibfield  {author} {\bibinfo {author} {\bibfnamefont {D.}~\bibnamefont {Guerci}}, \bibinfo {author} {\bibfnamefont {D.}~\bibnamefont {Kaplan}}, \bibinfo {author} {\bibfnamefont {J.}~\bibnamefont {Ingham}}, \bibinfo {author} {\bibfnamefont {J.~H.}\ \bibnamefont {Pixley}},\ and\ \bibinfo {author} {\bibfnamefont {A.~J.}\ \bibnamefont {Millis}},\ }\href@noop {} {\bibinfo {title} {Topological superconductivity from repulsive interactions in twisted {WSe$_2$}}} (\bibinfo {year} {2024}),\ \Eprint {https://arxiv.org/abs/2408.16075} {arXiv:2408.16075 [cond-mat.supr-con]} \BibitemShut {NoStop}%
\bibitem [{\citenamefont {Tokura}\ and\ \citenamefont {Nagaosa}(2018)}]{Tokura2018-yp}%
  \BibitemOpen
  \bibfield  {author} {\bibinfo {author} {\bibfnamefont {Y.}~\bibnamefont {Tokura}}\ and\ \bibinfo {author} {\bibfnamefont {N.}~\bibnamefont {Nagaosa}},\ }\bibfield  {title} {\bibinfo {title} {Nonreciprocal responses from non-centrosymmetric quantum materials},\ }\href {https://doi.org/10.1038/s41467-018-05759-4} {\bibfield  {journal} {\bibinfo  {journal} {Nat. Commun.}\ }\textbf {\bibinfo {volume} {9}},\ \bibinfo {pages} {3740} (\bibinfo {year} {2018})}\BibitemShut {NoStop}%
\bibitem [{\citenamefont {Nagaosa}\ and\ \citenamefont {Yanase}(2024)}]{Nagaosa2024-uk}%
  \BibitemOpen
  \bibfield  {author} {\bibinfo {author} {\bibfnamefont {N.}~\bibnamefont {Nagaosa}}\ and\ \bibinfo {author} {\bibfnamefont {Y.}~\bibnamefont {Yanase}},\ }\bibfield  {title} {\bibinfo {title} {{Nonreciprocal Transport and Optical Phenomena in Quantum Materials}},\ }\href {https://doi.org/10.1146/annurev-conmatphys-032822-033734} {\bibfield  {journal} {\bibinfo  {journal} {Annu. Rev. Condens. Matter Phys.}\ }\textbf {\bibinfo {volume} {15}},\ \bibinfo {pages} {63} (\bibinfo {year} {2024})}\BibitemShut {NoStop}%
\bibitem [{\citenamefont {Wang}\ \emph {et~al.}(2021)\citenamefont {Wang}, \citenamefont {Dong}, \citenamefont {Xiao},\ and\ \citenamefont {Niu}}]{Wang2021-bq}%
  \BibitemOpen
  \bibfield  {author} {\bibinfo {author} {\bibfnamefont {Z.}~\bibnamefont {Wang}}, \bibinfo {author} {\bibfnamefont {L.}~\bibnamefont {Dong}}, \bibinfo {author} {\bibfnamefont {C.}~\bibnamefont {Xiao}},\ and\ \bibinfo {author} {\bibfnamefont {Q.}~\bibnamefont {Niu}},\ }\bibfield  {title} {\bibinfo {title} {{Berry Curvature Effects on Quasiparticle Dynamics in Superconductors}},\ }\href {https://doi.org/10.1103/PhysRevLett.126.187001} {\bibfield  {journal} {\bibinfo  {journal} {Phys. Rev. Lett.}\ }\textbf {\bibinfo {volume} {126}},\ \bibinfo {pages} {187001} (\bibinfo {year} {2021})}\BibitemShut {NoStop}%
\bibitem [{\citenamefont {Watanabe}\ \emph {et~al.}(2022{\natexlab{a}})\citenamefont {Watanabe}, \citenamefont {Daido},\ and\ \citenamefont {Yanase}}]{Watanabe2022-pk}%
  \BibitemOpen
  \bibfield  {author} {\bibinfo {author} {\bibfnamefont {H.}~\bibnamefont {Watanabe}}, \bibinfo {author} {\bibfnamefont {A.}~\bibnamefont {Daido}},\ and\ \bibinfo {author} {\bibfnamefont {Y.}~\bibnamefont {Yanase}},\ }\bibfield  {title} {\bibinfo {title} {Nonreciprocal optical response in parity-breaking superconductors},\ }\href {https://doi.org/10.1103/physrevb.105.024308} {\bibfield  {journal} {\bibinfo  {journal} {Phys. Rev. B.}\ }\textbf {\bibinfo {volume} {105}},\ \bibinfo {pages} {024308} (\bibinfo {year} {2022}{\natexlab{a}})}\BibitemShut {NoStop}%
\bibitem [{\citenamefont {Watanabe}\ \emph {et~al.}(2022{\natexlab{b}})\citenamefont {Watanabe}, \citenamefont {Daido},\ and\ \citenamefont {Yanase}}]{Watanabe2022-af}%
  \BibitemOpen
  \bibfield  {author} {\bibinfo {author} {\bibfnamefont {H.}~\bibnamefont {Watanabe}}, \bibinfo {author} {\bibfnamefont {A.}~\bibnamefont {Daido}},\ and\ \bibinfo {author} {\bibfnamefont {Y.}~\bibnamefont {Yanase}},\ }\bibfield  {title} {\bibinfo {title} {{Nonreciprocal Meissner response in parity-mixed superconductors}},\ }\href {https://doi.org/10.1103/physrevb.105.l100504} {\bibfield  {journal} {\bibinfo  {journal} {Phys. Rev. B.}\ }\textbf {\bibinfo {volume} {105}},\ \bibinfo {pages} {L100504} (\bibinfo {year} {2022}{\natexlab{b}})}\BibitemShut {NoStop}%
\bibitem [{\citenamefont {Tanaka}\ \emph {et~al.}(2023)\citenamefont {Tanaka}, \citenamefont {Watanabe},\ and\ \citenamefont {Yanase}}]{Tanaka2023-st}%
  \BibitemOpen
  \bibfield  {author} {\bibinfo {author} {\bibfnamefont {H.}~\bibnamefont {Tanaka}}, \bibinfo {author} {\bibfnamefont {H.}~\bibnamefont {Watanabe}},\ and\ \bibinfo {author} {\bibfnamefont {Y.}~\bibnamefont {Yanase}},\ }\bibfield  {title} {\bibinfo {title} {Nonlinear optical responses in noncentrosymmetric superconductors},\ }\href {https://doi.org/10.1103/PhysRevB.107.024513} {\bibfield  {journal} {\bibinfo  {journal} {Phys. Rev. B}\ }\textbf {\bibinfo {volume} {107}},\ \bibinfo {pages} {024513} (\bibinfo {year} {2023})}\BibitemShut {NoStop}%
\bibitem [{\citenamefont {Tanaka}\ \emph {et~al.}(2024)\citenamefont {Tanaka}, \citenamefont {Watanabe},\ and\ \citenamefont {Yanase}}]{Tanaka2024-rt}%
  \BibitemOpen
  \bibfield  {author} {\bibinfo {author} {\bibfnamefont {H.}~\bibnamefont {Tanaka}}, \bibinfo {author} {\bibfnamefont {H.}~\bibnamefont {Watanabe}},\ and\ \bibinfo {author} {\bibfnamefont {Y.}~\bibnamefont {Yanase}},\ }\bibfield  {title} {\bibinfo {title} {Nonlinear optical response in superconductors in magnetic field: Quantum geometry and topological superconductivity},\ }\href {https://doi.org/10.1103/physrevb.110.014520} {\bibfield  {journal} {\bibinfo  {journal} {Phys. Rev. B.}\ }\textbf {\bibinfo {volume} {110}},\ \bibinfo {pages} {014520} (\bibinfo {year} {2024})}\BibitemShut {NoStop}%
\bibitem [{\citenamefont {Tanaka}\ and\ \citenamefont {Yanase}(2025)}]{Tanaka2025-dg}%
  \BibitemOpen
  \bibfield  {author} {\bibinfo {author} {\bibfnamefont {H.}~\bibnamefont {Tanaka}}\ and\ \bibinfo {author} {\bibfnamefont {Y.}~\bibnamefont {Yanase}},\ }\bibfield  {title} {\bibinfo {title} {Vertex correction for the linear and nonlinear optical responses in superconductors: Multiband effect and topological superconductivity},\ }\href {https://doi.org/10.1103/76lk-kj4x} {\bibfield  {journal} {\bibinfo  {journal} {Phys. Rev. B.}\ }\textbf {\bibinfo {volume} {112}},\ \bibinfo {pages} {094503} (\bibinfo {year} {2025})}\BibitemShut {NoStop}%
\bibitem [{\citenamefont {Kaplan}\ \emph {et~al.}(2025)\citenamefont {Kaplan}, \citenamefont {Lucht}, \citenamefont {Volkov},\ and\ \citenamefont {Pixley}}]{Kaplan2025-mz}%
  \BibitemOpen
  \bibfield  {author} {\bibinfo {author} {\bibfnamefont {D.}~\bibnamefont {Kaplan}}, \bibinfo {author} {\bibfnamefont {K.~P.}\ \bibnamefont {Lucht}}, \bibinfo {author} {\bibfnamefont {P.~A.}\ \bibnamefont {Volkov}},\ and\ \bibinfo {author} {\bibfnamefont {J.~H.}\ \bibnamefont {Pixley}},\ }\href@noop {} {\bibinfo {title} {Quantum geometric photocurrents of quasiparticles in superconductors}} (\bibinfo {year} {2025}),\ \Eprint {https://arxiv.org/abs/2502.12265} {arXiv:2502.12265 [cond-mat.supr-con]} \BibitemShut {NoStop}%
\bibitem [{\citenamefont {Matsyshyn}\ \emph {et~al.}(2026)\citenamefont {Matsyshyn}, \citenamefont {Vignale},\ and\ \citenamefont {Song}}]{Matsyshyn2026-ol}%
  \BibitemOpen
  \bibfield  {author} {\bibinfo {author} {\bibfnamefont {O.}~\bibnamefont {Matsyshyn}}, \bibinfo {author} {\bibfnamefont {G.}~\bibnamefont {Vignale}},\ and\ \bibinfo {author} {\bibfnamefont {J.~C.~W.}\ \bibnamefont {Song}},\ }\bibfield  {title} {\bibinfo {title} {{Superconducting Berry Curvature Dipole}},\ }\href {https://doi.org/10.1103/gbcm-l2qd} {\bibfield  {journal} {\bibinfo  {journal} {Phys. Rev. Lett.}\ }\textbf {\bibinfo {volume} {136}},\ \bibinfo {pages} {246902} (\bibinfo {year} {2026})}\BibitemShut {NoStop}%
\bibitem [{\citenamefont {Fal'ko}\ \emph {et~al.}(1993)\citenamefont {Fal'ko}, \citenamefont {Meshkov},\ and\ \citenamefont {Iordanskii}}]{Fal-ko1993-dz}%
  \BibitemOpen
  \bibfield  {author} {\bibinfo {author} {\bibfnamefont {V.~I.}\ \bibnamefont {Fal'ko}, \bibfnamefont {VI}}, \bibinfo {author} {\bibfnamefont {S.~V.}\ \bibnamefont {Meshkov}},\ and\ \bibinfo {author} {\bibfnamefont {S.~V.}\ \bibnamefont {Iordanskii}},\ }\bibfield  {title} {\bibinfo {title} {Acoustoelectric drag effect in the two-dimensional electron gas at strong magnetic field},\ }\href {https://doi.org/10.1103/physrevb.47.9910} {\bibfield  {journal} {\bibinfo  {journal} {Phys. Rev. B}\ }\textbf {\bibinfo {volume} {47}},\ \bibinfo {pages} {9910} (\bibinfo {year} {1993})}\BibitemShut {NoStop}%
\bibitem [{\citenamefont {Miseikis}\ \emph {et~al.}(2012)\citenamefont {Miseikis}, \citenamefont {Cunningham}, \citenamefont {Saeed}, \citenamefont {O'Rorke},\ and\ \citenamefont {Davies}}]{Miseikis2012-zp}%
  \BibitemOpen
  \bibfield  {author} {\bibinfo {author} {\bibfnamefont {V.}~\bibnamefont {Miseikis}}, \bibinfo {author} {\bibfnamefont {J.~E.}\ \bibnamefont {Cunningham}}, \bibinfo {author} {\bibfnamefont {K.}~\bibnamefont {Saeed}}, \bibinfo {author} {\bibfnamefont {R.}~\bibnamefont {O'Rorke}},\ and\ \bibinfo {author} {\bibfnamefont {A.~G.}\ \bibnamefont {Davies}},\ }\bibfield  {title} {\bibinfo {title} {Acoustically induced current flow in graphene},\ }\href {https://doi.org/10.1063/1.3697403} {\bibfield  {journal} {\bibinfo  {journal} {Appl. Phys. Lett.}\ }\textbf {\bibinfo {volume} {100}},\ \bibinfo {pages} {133105} (\bibinfo {year} {2012})}\BibitemShut {NoStop}%
\bibitem [{\citenamefont {Bandhu}\ \emph {et~al.}(2013)\citenamefont {Bandhu}, \citenamefont {Lawton},\ and\ \citenamefont {Nash}}]{Bandhu2013-qq}%
  \BibitemOpen
  \bibfield  {author} {\bibinfo {author} {\bibfnamefont {L.}~\bibnamefont {Bandhu}}, \bibinfo {author} {\bibfnamefont {L.~M.}\ \bibnamefont {Lawton}},\ and\ \bibinfo {author} {\bibfnamefont {G.~R.}\ \bibnamefont {Nash}},\ }\bibfield  {title} {\bibinfo {title} {Macroscopic acoustoelectric charge transport in graphene},\ }\href {https://doi.org/10.1063/1.4822121} {\bibfield  {journal} {\bibinfo  {journal} {Appl. Phys. Lett.}\ }\textbf {\bibinfo {volume} {103}},\ \bibinfo {pages} {133101} (\bibinfo {year} {2013})}\BibitemShut {NoStop}%
\bibitem [{\citenamefont {Sonowal}\ \emph {et~al.}(2020)\citenamefont {Sonowal}, \citenamefont {Kalameitsev}, \citenamefont {Kovalev},\ and\ \citenamefont {Savenko}}]{Sonowal2020-bt}%
  \BibitemOpen
  \bibfield  {author} {\bibinfo {author} {\bibfnamefont {K.}~\bibnamefont {Sonowal}}, \bibinfo {author} {\bibfnamefont {A.~V.}\ \bibnamefont {Kalameitsev}}, \bibinfo {author} {\bibfnamefont {V.~M.}\ \bibnamefont {Kovalev}},\ and\ \bibinfo {author} {\bibfnamefont {I.~G.}\ \bibnamefont {Savenko}},\ }\bibfield  {title} {\bibinfo {title} {{Acoustoelectric effect in two-dimensional Dirac materials exposed to Rayleigh surface acoustic waves}},\ }\href {https://doi.org/10.1103/physrevb.102.235405} {\bibfield  {journal} {\bibinfo  {journal} {Phys. Rev. B.}\ }\textbf {\bibinfo {volume} {102}},\ \bibinfo {pages} {235405} (\bibinfo {year} {2020})}\BibitemShut {NoStop}%
\bibitem [{\citenamefont {Mou}\ \emph {et~al.}(2025)\citenamefont {Mou}, \citenamefont {Wang}, \citenamefont {Chen}, \citenamefont {Xia}, \citenamefont {Li}, \citenamefont {Yan}, \citenamefont {Jiang}, \citenamefont {Wu}, \citenamefont {Shi}, \citenamefont {Jiang}, \citenamefont {Xie},\ and\ \citenamefont {Zhang}}]{Mou2025-dm}%
  \BibitemOpen
  \bibfield  {author} {\bibinfo {author} {\bibfnamefont {Y.}~\bibnamefont {Mou}}, \bibinfo {author} {\bibfnamefont {J.}~\bibnamefont {Wang}}, \bibinfo {author} {\bibfnamefont {H.}~\bibnamefont {Chen}}, \bibinfo {author} {\bibfnamefont {Y.}~\bibnamefont {Xia}}, \bibinfo {author} {\bibfnamefont {H.}~\bibnamefont {Li}}, \bibinfo {author} {\bibfnamefont {Q.}~\bibnamefont {Yan}}, \bibinfo {author} {\bibfnamefont {X.}~\bibnamefont {Jiang}}, \bibinfo {author} {\bibfnamefont {Y.}~\bibnamefont {Wu}}, \bibinfo {author} {\bibfnamefont {W.}~\bibnamefont {Shi}}, \bibinfo {author} {\bibfnamefont {H.}~\bibnamefont {Jiang}}, \bibinfo {author} {\bibfnamefont {X.~C.}\ \bibnamefont {Xie}},\ and\ \bibinfo {author} {\bibfnamefont {C.}~\bibnamefont {Zhang}},\ }\bibfield  {title} {\bibinfo {title} {{Coherent Detection of the Oscillating Acoustoelectric Effect in Graphene}},\ }\href {https://doi.org/10.1103/PhysRevLett.134.096301} {\bibfield  {journal} {\bibinfo  {journal} {Phys. Rev. Lett.}\ }\textbf {\bibinfo {volume} {134}},\
  \bibinfo {pages} {096301} (\bibinfo {year} {2025})}\BibitemShut {NoStop}%
\bibitem [{\citenamefont {von Oppen}\ \emph {et~al.}(2009)\citenamefont {von Oppen}, \citenamefont {Guinea},\ and\ \citenamefont {Mariani}}]{von-Oppen2009-at}%
  \BibitemOpen
  \bibfield  {author} {\bibinfo {author} {\bibfnamefont {F.}~\bibnamefont {von Oppen}}, \bibinfo {author} {\bibfnamefont {F.}~\bibnamefont {Guinea}},\ and\ \bibinfo {author} {\bibfnamefont {E.}~\bibnamefont {Mariani}},\ }\bibfield  {title} {\bibinfo {title} {Synthetic electric fields and phonon damping in carbon nanotubes and graphene},\ }\href {https://doi.org/10.1103/PhysRevB.80.075420} {\bibfield  {journal} {\bibinfo  {journal} {Phys. Rev. B}\ }\textbf {\bibinfo {volume} {80}},\ \bibinfo {pages} {075420} (\bibinfo {year} {2009})}\BibitemShut {NoStop}%
\bibitem [{\citenamefont {Vaezi}\ \emph {et~al.}(2013)\citenamefont {Vaezi}, \citenamefont {Abedpour}, \citenamefont {Asgari}, \citenamefont {Cortijo},\ and\ \citenamefont {Vozmediano}}]{Vaezi2013-uc}%
  \BibitemOpen
  \bibfield  {author} {\bibinfo {author} {\bibfnamefont {A.}~\bibnamefont {Vaezi}}, \bibinfo {author} {\bibfnamefont {N.}~\bibnamefont {Abedpour}}, \bibinfo {author} {\bibfnamefont {R.}~\bibnamefont {Asgari}}, \bibinfo {author} {\bibfnamefont {A.}~\bibnamefont {Cortijo}},\ and\ \bibinfo {author} {\bibfnamefont {M.~A.~H.}\ \bibnamefont {Vozmediano}},\ }\bibfield  {title} {\bibinfo {title} {Topological electric current from time-dependent elastic deformations in graphene},\ }\href {https://doi.org/10.1103/PhysRevB.88.125406} {\bibfield  {journal} {\bibinfo  {journal} {Phys. Rev. B}\ }\textbf {\bibinfo {volume} {88}},\ \bibinfo {pages} {125406} (\bibinfo {year} {2013})}\BibitemShut {NoStop}%
\bibitem [{\citenamefont {Cazalilla}\ \emph {et~al.}(2014)\citenamefont {Cazalilla}, \citenamefont {Ochoa},\ and\ \citenamefont {Guinea}}]{Cazalilla2014-eh}%
  \BibitemOpen
  \bibfield  {author} {\bibinfo {author} {\bibfnamefont {M.~A.}\ \bibnamefont {Cazalilla}}, \bibinfo {author} {\bibfnamefont {H.}~\bibnamefont {Ochoa}},\ and\ \bibinfo {author} {\bibfnamefont {F.}~\bibnamefont {Guinea}},\ }\bibfield  {title} {\bibinfo {title} {{Quantum Spin Hall Effect in Two-Dimensional Crystals of Transition-Metal Dichalcogenides}},\ }\href {https://doi.org/10.1103/PhysRevLett.113.077201} {\bibfield  {journal} {\bibinfo  {journal} {Phys. Rev. Lett.}\ }\textbf {\bibinfo {volume} {113}},\ \bibinfo {pages} {077201} (\bibinfo {year} {2014})}\BibitemShut {NoStop}%
\bibitem [{\citenamefont {Kalameitsev}\ \emph {et~al.}(2019)\citenamefont {Kalameitsev}, \citenamefont {Kovalev},\ and\ \citenamefont {Savenko}}]{Kalameitsev2019-os}%
  \BibitemOpen
  \bibfield  {author} {\bibinfo {author} {\bibfnamefont {A.~V.}\ \bibnamefont {Kalameitsev}}, \bibinfo {author} {\bibfnamefont {V.~M.}\ \bibnamefont {Kovalev}},\ and\ \bibinfo {author} {\bibfnamefont {I.~G.}\ \bibnamefont {Savenko}},\ }\bibfield  {title} {\bibinfo {title} {{Valley Acoustoelectric Effect}},\ }\href {https://doi.org/10.1103/physrevlett.122.256801} {\bibfield  {journal} {\bibinfo  {journal} {Phys. Rev. Lett.}\ }\textbf {\bibinfo {volume} {122}},\ \bibinfo {pages} {256801} (\bibinfo {year} {2019})}\BibitemShut {NoStop}%
\bibitem [{\citenamefont {Sukhachov}\ and\ \citenamefont {Rostami}(2020)}]{Sukhachov2020-ri}%
  \BibitemOpen
  \bibfield  {author} {\bibinfo {author} {\bibfnamefont {P.~O.}\ \bibnamefont {Sukhachov}}\ and\ \bibinfo {author} {\bibfnamefont {H.}~\bibnamefont {Rostami}},\ }\bibfield  {title} {\bibinfo {title} {{Acoustogalvanic Effect in Dirac and Weyl Semimetals}},\ }\href {https://doi.org/10.1103/PhysRevLett.124.126602} {\bibfield  {journal} {\bibinfo  {journal} {Phys. Rev. Lett.}\ }\textbf {\bibinfo {volume} {124}},\ \bibinfo {pages} {126602} (\bibinfo {year} {2020})}\BibitemShut {NoStop}%
\bibitem [{\citenamefont {Sela}\ \emph {et~al.}(2020)\citenamefont {Sela}, \citenamefont {Bloch}, \citenamefont {von Oppen},\ and\ \citenamefont {Shalom}}]{Sela2020-gl}%
  \BibitemOpen
  \bibfield  {author} {\bibinfo {author} {\bibfnamefont {E.}~\bibnamefont {Sela}}, \bibinfo {author} {\bibfnamefont {Y.}~\bibnamefont {Bloch}}, \bibinfo {author} {\bibfnamefont {F.}~\bibnamefont {von Oppen}},\ and\ \bibinfo {author} {\bibfnamefont {M.~B.}\ \bibnamefont {Shalom}},\ }\bibfield  {title} {\bibinfo {title} {{Quantum Hall Response to Time-Dependent Strain Gradients in Graphene}},\ }\href {https://doi.org/10.1103/PhysRevLett.124.026602} {\bibfield  {journal} {\bibinfo  {journal} {Phys. Rev. Lett.}\ }\textbf {\bibinfo {volume} {124}},\ \bibinfo {pages} {026602} (\bibinfo {year} {2020})}\BibitemShut {NoStop}%
\bibitem [{\citenamefont {Li}\ \emph {et~al.}(2020)\citenamefont {Li}, \citenamefont {Su}, \citenamefont {Ren},\ and\ \citenamefont {He}}]{Li2020-ch}%
  \BibitemOpen
  \bibfield  {author} {\bibinfo {author} {\bibfnamefont {S.-Y.}\ \bibnamefont {Li}}, \bibinfo {author} {\bibfnamefont {Y.}~\bibnamefont {Su}}, \bibinfo {author} {\bibfnamefont {Y.-N.}\ \bibnamefont {Ren}},\ and\ \bibinfo {author} {\bibfnamefont {L.}~\bibnamefont {He}},\ }\bibfield  {title} {\bibinfo {title} {{Valley Polarization and Inversion in Strained Graphene via Pseudo-Landau Levels, Valley Splitting of Real Landau Levels, and Confined States}},\ }\href {https://doi.org/10.1103/PhysRevLett.124.106802} {\bibfield  {journal} {\bibinfo  {journal} {Phys. Rev. Lett.}\ }\textbf {\bibinfo {volume} {124}},\ \bibinfo {pages} {106802} (\bibinfo {year} {2020})}\BibitemShut {NoStop}%
\bibitem [{\citenamefont {Zhao}\ \emph {et~al.}(2022)\citenamefont {Zhao}, \citenamefont {Sharma}, \citenamefont {Liang}, \citenamefont {Glasenapp}, \citenamefont {Mourokh}, \citenamefont {Kovalev}, \citenamefont {Huber}, \citenamefont {Prada}, \citenamefont {Tiemann},\ and\ \citenamefont {Blick}}]{Zhao2022-hr}%
  \BibitemOpen
  \bibfield  {author} {\bibinfo {author} {\bibfnamefont {P.}~\bibnamefont {Zhao}}, \bibinfo {author} {\bibfnamefont {C.~H.}\ \bibnamefont {Sharma}}, \bibinfo {author} {\bibfnamefont {R.}~\bibnamefont {Liang}}, \bibinfo {author} {\bibfnamefont {C.}~\bibnamefont {Glasenapp}}, \bibinfo {author} {\bibfnamefont {L.}~\bibnamefont {Mourokh}}, \bibinfo {author} {\bibfnamefont {V.~M.}\ \bibnamefont {Kovalev}}, \bibinfo {author} {\bibfnamefont {P.}~\bibnamefont {Huber}}, \bibinfo {author} {\bibfnamefont {M.}~\bibnamefont {Prada}}, \bibinfo {author} {\bibfnamefont {L.}~\bibnamefont {Tiemann}},\ and\ \bibinfo {author} {\bibfnamefont {R.~H.}\ \bibnamefont {Blick}},\ }\bibfield  {title} {\bibinfo {title} {{Acoustically Induced Giant Synthetic Hall Voltages in Graphene}},\ }\href {https://doi.org/10.1103/PhysRevLett.128.256601} {\bibfield  {journal} {\bibinfo  {journal} {Phys. Rev. Lett.}\ }\textbf {\bibinfo {volume} {128}},\ \bibinfo {pages} {256601} (\bibinfo {year} {2022})}\BibitemShut {NoStop}%
\bibitem [{\citenamefont {Bhalla}\ \emph {et~al.}(2022)\citenamefont {Bhalla}, \citenamefont {Vignale},\ and\ \citenamefont {Rostami}}]{Bhalla2022-su}%
  \BibitemOpen
  \bibfield  {author} {\bibinfo {author} {\bibfnamefont {P.}~\bibnamefont {Bhalla}}, \bibinfo {author} {\bibfnamefont {G.}~\bibnamefont {Vignale}},\ and\ \bibinfo {author} {\bibfnamefont {H.}~\bibnamefont {Rostami}},\ }\bibfield  {title} {\bibinfo {title} {{Pseudogauge field driven acoustoelectric current in two-dimensional hexagonal Dirac materials}},\ }\href {https://doi.org/10.1103/physrevb.105.125407} {\bibfield  {journal} {\bibinfo  {journal} {Phys. Rev. B.}\ }\textbf {\bibinfo {volume} {105}},\ \bibinfo {pages} {125407} (\bibinfo {year} {2022})}\BibitemShut {NoStop}%
\bibitem [{\citenamefont {Uchoa}\ and\ \citenamefont {Barlas}(2013)}]{Uchoa2013-yi}%
  \BibitemOpen
  \bibfield  {author} {\bibinfo {author} {\bibfnamefont {B.}~\bibnamefont {Uchoa}}\ and\ \bibinfo {author} {\bibfnamefont {Y.}~\bibnamefont {Barlas}},\ }\bibfield  {title} {\bibinfo {title} {{Superconducting States in Pseudo-Landau-Levels of Strained Graphene}},\ }\href {https://doi.org/10.1103/PhysRevLett.111.046604} {\bibfield  {journal} {\bibinfo  {journal} {Phys. Rev. Lett.}\ }\textbf {\bibinfo {volume} {111}},\ \bibinfo {pages} {046604} (\bibinfo {year} {2013})}\BibitemShut {NoStop}%
\bibitem [{\citenamefont {Massarelli}\ \emph {et~al.}(2017)\citenamefont {Massarelli}, \citenamefont {Wachtel}, \citenamefont {Wei},\ and\ \citenamefont {Paramekanti}}]{Massarelli2017-sc}%
  \BibitemOpen
  \bibfield  {author} {\bibinfo {author} {\bibfnamefont {G.}~\bibnamefont {Massarelli}}, \bibinfo {author} {\bibfnamefont {G.}~\bibnamefont {Wachtel}}, \bibinfo {author} {\bibfnamefont {J.~Y.~T.}\ \bibnamefont {Wei}},\ and\ \bibinfo {author} {\bibfnamefont {A.}~\bibnamefont {Paramekanti}},\ }\bibfield  {title} {\bibinfo {title} {{Pseudo-Landau levels of Bogoliubov quasiparticles in strained nodal superconductors}},\ }\href {https://doi.org/10.1103/PhysRevB.96.224516} {\bibfield  {journal} {\bibinfo  {journal} {Phys. Rev. B}\ }\textbf {\bibinfo {volume} {96}},\ \bibinfo {pages} {224516} (\bibinfo {year} {2017})}\BibitemShut {NoStop}%
\bibitem [{\citenamefont {Nica}\ and\ \citenamefont {Franz}(2018)}]{Nica2018-wy}%
  \BibitemOpen
  \bibfield  {author} {\bibinfo {author} {\bibfnamefont {E.~M.}\ \bibnamefont {Nica}}\ and\ \bibinfo {author} {\bibfnamefont {M.}~\bibnamefont {Franz}},\ }\bibfield  {title} {\bibinfo {title} {{Landau levels from neutral Bogoliubov particles in two-dimensional nodal superconductors under strain and doping gradients}},\ }\href {https://doi.org/10.1103/PhysRevB.97.024520} {\bibfield  {journal} {\bibinfo  {journal} {Phys. Rev. B}\ }\textbf {\bibinfo {volume} {97}},\ \bibinfo {pages} {024520} (\bibinfo {year} {2018})}\BibitemShut {NoStop}%
\bibitem [{\citenamefont {Nayga}\ \emph {et~al.}(2019)\citenamefont {Nayga}, \citenamefont {Rachel},\ and\ \citenamefont {Vojta}}]{Nayga2019-jx}%
  \BibitemOpen
  \bibfield  {author} {\bibinfo {author} {\bibfnamefont {M.~M.}\ \bibnamefont {Nayga}}, \bibinfo {author} {\bibfnamefont {S.}~\bibnamefont {Rachel}},\ and\ \bibinfo {author} {\bibfnamefont {M.}~\bibnamefont {Vojta}},\ }\bibfield  {title} {\bibinfo {title} {{Magnon Landau Levels and Emergent Supersymmetry in Strained Antiferromagnets}},\ }\href {https://doi.org/10.1103/PhysRevLett.123.207204} {\bibfield  {journal} {\bibinfo  {journal} {Phys. Rev. Lett.}\ }\textbf {\bibinfo {volume} {123}},\ \bibinfo {pages} {207204} (\bibinfo {year} {2019})}\BibitemShut {NoStop}%
\bibitem [{\citenamefont {Sano}\ \emph {et~al.}(2024)\citenamefont {Sano}, \citenamefont {Ominato},\ and\ \citenamefont {Matsuo}}]{Sano2024-gy}%
  \BibitemOpen
  \bibfield  {author} {\bibinfo {author} {\bibfnamefont {R.}~\bibnamefont {Sano}}, \bibinfo {author} {\bibfnamefont {Y.}~\bibnamefont {Ominato}},\ and\ \bibinfo {author} {\bibfnamefont {M.}~\bibnamefont {Matsuo}},\ }\bibfield  {title} {\bibinfo {title} {{Acoustomagnonic Spin Hall Effect in Honeycomb Antiferromagnets}},\ }\href {https://doi.org/10.1103/PhysRevLett.132.236302} {\bibfield  {journal} {\bibinfo  {journal} {Phys. Rev. Lett.}\ }\textbf {\bibinfo {volume} {132}},\ \bibinfo {pages} {236302} (\bibinfo {year} {2024})}\BibitemShut {NoStop}%
\bibitem [{\citenamefont {Yamazaki}\ \emph {et~al.}(2023)\citenamefont {Yamazaki}, \citenamefont {Funato},\ and\ \citenamefont {Yamakage}}]{Yamazaki2023-av}%
  \BibitemOpen
  \bibfield  {author} {\bibinfo {author} {\bibfnamefont {Y.}~\bibnamefont {Yamazaki}}, \bibinfo {author} {\bibfnamefont {T.}~\bibnamefont {Funato}},\ and\ \bibinfo {author} {\bibfnamefont {A.}~\bibnamefont {Yamakage}},\ }\bibfield  {title} {\bibinfo {title} {Majorana spin current generation by dynamic strain},\ }\href {https://doi.org/10.1103/physrevb.108.l060505} {\bibfield  {journal} {\bibinfo  {journal} {Phys. Rev. B.}\ }\textbf {\bibinfo {volume} {108}},\ \bibinfo {pages} {L060505} (\bibinfo {year} {2023})}\BibitemShut {NoStop}%
\bibitem [{\citenamefont {Wu}\ \emph {et~al.}(2019)\citenamefont {Wu}, \citenamefont {Lovorn}, \citenamefont {Tutuc}, \citenamefont {Martin},\ and\ \citenamefont {MacDonald}}]{Wu2019-ry}%
  \BibitemOpen
  \bibfield  {author} {\bibinfo {author} {\bibfnamefont {F.}~\bibnamefont {Wu}}, \bibinfo {author} {\bibfnamefont {T.}~\bibnamefont {Lovorn}}, \bibinfo {author} {\bibfnamefont {E.}~\bibnamefont {Tutuc}}, \bibinfo {author} {\bibfnamefont {I.}~\bibnamefont {Martin}},\ and\ \bibinfo {author} {\bibfnamefont {A.~H.}\ \bibnamefont {MacDonald}},\ }\bibfield  {title} {\bibinfo {title} {{Topological Insulators in Twisted Transition Metal Dichalcogenide Homobilayers}},\ }\href {https://doi.org/10.1103/PhysRevLett.122.086402} {\bibfield  {journal} {\bibinfo  {journal} {Phys. Rev. Lett.}\ }\textbf {\bibinfo {volume} {122}},\ \bibinfo {pages} {086402} (\bibinfo {year} {2019})}\BibitemShut {NoStop}%
\bibitem [{\citenamefont {Devakul}\ \emph {et~al.}(2021)\citenamefont {Devakul}, \citenamefont {Crépel}, \citenamefont {Zhang},\ and\ \citenamefont {Fu}}]{Devakul2021-hs}%
  \BibitemOpen
  \bibfield  {author} {\bibinfo {author} {\bibfnamefont {T.}~\bibnamefont {Devakul}}, \bibinfo {author} {\bibfnamefont {V.}~\bibnamefont {Crépel}}, \bibinfo {author} {\bibfnamefont {Y.}~\bibnamefont {Zhang}},\ and\ \bibinfo {author} {\bibfnamefont {L.}~\bibnamefont {Fu}},\ }\bibfield  {title} {\bibinfo {title} {Magic in twisted transition metal dichalcogenide bilayers},\ }\href {https://doi.org/10.1038/s41467-021-27042-9} {\bibfield  {journal} {\bibinfo  {journal} {Nat. Commun.}\ }\textbf {\bibinfo {volume} {12}},\ \bibinfo {pages} {6730} (\bibinfo {year} {2021})}\BibitemShut {NoStop}%
\bibitem [{\citenamefont {Zhu}\ \emph {et~al.}(2025)\citenamefont {Zhu}, \citenamefont {Chou}, \citenamefont {Xie},\ and\ \citenamefont {Das~Sarma}}]{Zhu2025-ba}%
  \BibitemOpen
  \bibfield  {author} {\bibinfo {author} {\bibfnamefont {J.}~\bibnamefont {Zhu}}, \bibinfo {author} {\bibfnamefont {Y.-Z.}\ \bibnamefont {Chou}}, \bibinfo {author} {\bibfnamefont {M.}~\bibnamefont {Xie}},\ and\ \bibinfo {author} {\bibfnamefont {S.}~\bibnamefont {Das~Sarma}},\ }\bibfield  {title} {\bibinfo {title} {Superconductivity in twisted transition metal dichalcogenide homobilayers},\ }\href {https://doi.org/10.1103/physrevb.111.l060501} {\bibfield  {journal} {\bibinfo  {journal} {Phys. Rev. B.}\ }\textbf {\bibinfo {volume} {111}},\ \bibinfo {pages} {L060501} (\bibinfo {year} {2025})}\BibitemShut {NoStop}%
\bibitem [{End()}]{Endmatter}%
  \BibitemOpen
  \href@noop {} {}\bibinfo {note} {See End Matter for more details.}\BibitemShut {Stop}%
\bibitem [{Sup()}]{Supplemental}%
  \BibitemOpen
  \href@noop {} {}\bibinfo {note} {See Supplemental Material for more details.}\BibitemShut {Stop}%
\bibitem [{\citenamefont {Nie}\ \emph {et~al.}(2023)\citenamefont {Nie}, \citenamefont {Wu}, \citenamefont {Wang}, \citenamefont {Ban}, \citenamefont {Lei}, \citenamefont {Yi}, \citenamefont {Liu},\ and\ \citenamefont {Liu}}]{Nie2023-cd}%
  \BibitemOpen
  \bibfield  {author} {\bibinfo {author} {\bibfnamefont {X.}~\bibnamefont {Nie}}, \bibinfo {author} {\bibfnamefont {X.}~\bibnamefont {Wu}}, \bibinfo {author} {\bibfnamefont {Y.}~\bibnamefont {Wang}}, \bibinfo {author} {\bibfnamefont {S.}~\bibnamefont {Ban}}, \bibinfo {author} {\bibfnamefont {Z.}~\bibnamefont {Lei}}, \bibinfo {author} {\bibfnamefont {J.}~\bibnamefont {Yi}}, \bibinfo {author} {\bibfnamefont {Y.}~\bibnamefont {Liu}},\ and\ \bibinfo {author} {\bibfnamefont {Y.}~\bibnamefont {Liu}},\ }\bibfield  {title} {\bibinfo {title} {Surface acoustic wave induced phenomena in two-dimensional materials},\ }\href {https://doi.org/10.1039/d2nh00458e} {\bibfield  {journal} {\bibinfo  {journal} {Nanoscale Horiz}\ }\textbf {\bibinfo {volume} {8}},\ \bibinfo {pages} {158} (\bibinfo {year} {2023})}\BibitemShut {NoStop}%
\bibitem [{\citenamefont {Landau}\ and\ \citenamefont {Lifshitz}(1986)}]{Landau1986-fa}%
  \BibitemOpen
  \bibfield  {author} {\bibinfo {author} {\bibfnamefont {L.~D.}\ \bibnamefont {Landau}}\ and\ \bibinfo {author} {\bibfnamefont {E.~M.}\ \bibnamefont {Lifshitz}},\ }\href@noop {} {\emph {\bibinfo {title} {Theory of Elasticity}}},\ \bibinfo {edition} {3rd}\ ed.,\ \bibinfo {series} {Course of Theoretical Physics}, Vol.~\bibinfo {volume} {7}\ (\bibinfo  {publisher} {Pergamon Press},\ \bibinfo {address} {Oxford},\ \bibinfo {year} {1986})\BibitemShut {NoStop}%
\bibitem [{\citenamefont {Rostami}\ \emph {et~al.}(2015)\citenamefont {Rostami}, \citenamefont {Rold\'{a}n}, \citenamefont {Cappelluti}, \citenamefont {Asgari},\ and\ \citenamefont {Guinea}}]{Rostami2015-hh}%
  \BibitemOpen
  \bibfield  {author} {\bibinfo {author} {\bibfnamefont {H.}~\bibnamefont {Rostami}}, \bibinfo {author} {\bibfnamefont {R.}~\bibnamefont {Rold\'{a}n}}, \bibinfo {author} {\bibfnamefont {E.}~\bibnamefont {Cappelluti}}, \bibinfo {author} {\bibfnamefont {R.}~\bibnamefont {Asgari}},\ and\ \bibinfo {author} {\bibfnamefont {F.}~\bibnamefont {Guinea}},\ }\bibfield  {title} {\bibinfo {title} {{Theory of strain in single-layer transition metal dichalcogenides}},\ }\href {https://doi.org/10.1103/physrevb.92.195402} {\bibfield  {journal} {\bibinfo  {journal} {Phys. Rev. B.}\ }\textbf {\bibinfo {volume} {92}},\ \bibinfo {pages} {195402} (\bibinfo {year} {2015})}\BibitemShut {NoStop}%
\bibitem [{\citenamefont {Hu}\ \emph {et~al.}(2022)\citenamefont {Hu}, \citenamefont {Zhang}, \citenamefont {Xie},\ and\ \citenamefont {Law}}]{Hu2022-im}%
  \BibitemOpen
  \bibfield  {author} {\bibinfo {author} {\bibfnamefont {J.-X.}\ \bibnamefont {Hu}}, \bibinfo {author} {\bibfnamefont {C.-P.}\ \bibnamefont {Zhang}}, \bibinfo {author} {\bibfnamefont {Y.-M.}\ \bibnamefont {Xie}},\ and\ \bibinfo {author} {\bibfnamefont {K.~T.}\ \bibnamefont {Law}},\ }\bibfield  {title} {\bibinfo {title} {{Nonlinear Hall effects in strained twisted bilayer {WSe$_2$}}},\ }\href {https://doi.org/10.1038/s42005-022-01034-7} {\bibfield  {journal} {\bibinfo  {journal} {Commun. Phys.}\ }\textbf {\bibinfo {volume} {5}},\ \bibinfo {pages} {1} (\bibinfo {year} {2022})}\BibitemShut {NoStop}%
\bibitem [{\citenamefont {Ahn}\ \emph {et~al.}(2020)\citenamefont {Ahn}, \citenamefont {Guo},\ and\ \citenamefont {Nagaosa}}]{Ahn2020-ee}%
  \BibitemOpen
  \bibfield  {author} {\bibinfo {author} {\bibfnamefont {J.}~\bibnamefont {Ahn}}, \bibinfo {author} {\bibfnamefont {G.-Y.}\ \bibnamefont {Guo}},\ and\ \bibinfo {author} {\bibfnamefont {N.}~\bibnamefont {Nagaosa}},\ }\bibfield  {title} {\bibinfo {title} {{Low-Frequency Divergence and Quantum Geometry of the Bulk Photovoltaic Effect in Topological Semimetals}},\ }\href {https://doi.org/10.1103/physrevx.10.041041} {\bibfield  {journal} {\bibinfo  {journal} {Phys. Rev. X.}\ }\textbf {\bibinfo {volume} {10}},\ \bibinfo {pages} {041041} (\bibinfo {year} {2020})}\BibitemShut {NoStop}%
\bibitem [{\citenamefont {Watanabe}\ and\ \citenamefont {Yanase}(2021)}]{Watanabe2021-cf}%
  \BibitemOpen
  \bibfield  {author} {\bibinfo {author} {\bibfnamefont {H.}~\bibnamefont {Watanabe}}\ and\ \bibinfo {author} {\bibfnamefont {Y.}~\bibnamefont {Yanase}},\ }\bibfield  {title} {\bibinfo {title} {{Chiral Photocurrent in Parity-Violating Magnet and Enhanced Response in Topological Antiferromagnet}},\ }\href {https://doi.org/10.1103/physrevx.11.011001} {\bibfield  {journal} {\bibinfo  {journal} {Phys. Rev. X.}\ }\textbf {\bibinfo {volume} {11}},\ \bibinfo {pages} {011001} (\bibinfo {year} {2021})}\BibitemShut {NoStop}%
\bibitem [{\citenamefont {Fang}\ \emph {et~al.}(2018)\citenamefont {Fang}, \citenamefont {Carr}, \citenamefont {Cazalilla},\ and\ \citenamefont {Kaxiras}}]{Fang2018-wv}%
  \BibitemOpen
  \bibfield  {author} {\bibinfo {author} {\bibfnamefont {S.}~\bibnamefont {Fang}}, \bibinfo {author} {\bibfnamefont {S.}~\bibnamefont {Carr}}, \bibinfo {author} {\bibfnamefont {M.~A.}\ \bibnamefont {Cazalilla}},\ and\ \bibinfo {author} {\bibfnamefont {E.}~\bibnamefont {Kaxiras}},\ }\bibfield  {title} {\bibinfo {title} {Electronic structure theory of strained two-dimensional materials with hexagonal symmetry},\ }\href {https://doi.org/10.1103/physrevb.98.075106} {\bibfield  {journal} {\bibinfo  {journal} {Phys. Rev. B.}\ }\textbf {\bibinfo {volume} {98}},\ \bibinfo {pages} {075106} (\bibinfo {year} {2018})}\BibitemShut {NoStop}%
\bibitem [{\citenamefont {Bi}\ \emph {et~al.}(2019)\citenamefont {Bi}, \citenamefont {Yuan},\ and\ \citenamefont {Fu}}]{Bi2019-aa}%
  \BibitemOpen
  \bibfield  {author} {\bibinfo {author} {\bibfnamefont {Z.}~\bibnamefont {Bi}}, \bibinfo {author} {\bibfnamefont {N.~F.~Q.}\ \bibnamefont {Yuan}},\ and\ \bibinfo {author} {\bibfnamefont {L.}~\bibnamefont {Fu}},\ }\bibfield  {title} {\bibinfo {title} {Designing flat bands by strain},\ }\href {https://doi.org/10.1103/PhysRevB.100.035448} {\bibfield  {journal} {\bibinfo  {journal} {Phys. Rev. B}\ }\textbf {\bibinfo {volume} {100}},\ \bibinfo {pages} {035448} (\bibinfo {year} {2019})}\BibitemShut {NoStop}%
\bibitem [{\citenamefont {Maznev}\ \emph {et~al.}(2021)\citenamefont {Maznev}, \citenamefont {Mincigrucci}, \citenamefont {Bencivenga}, \citenamefont {Unikandanunni}, \citenamefont {Capotondi}, \citenamefont {Chen}, \citenamefont {Ding}, \citenamefont {Duncan}, \citenamefont {Foglia}, \citenamefont {Izzo}, \citenamefont {Masciovecchio}, \citenamefont {Martinelli}, \citenamefont {Monaco}, \citenamefont {Pedersoli}, \citenamefont {Bonetti},\ and\ \citenamefont {Nelson}}]{Maznev2021-ou}%
  \BibitemOpen
  \bibfield  {author} {\bibinfo {author} {\bibfnamefont {A.~A.}\ \bibnamefont {Maznev}}, \bibinfo {author} {\bibfnamefont {R.}~\bibnamefont {Mincigrucci}}, \bibinfo {author} {\bibfnamefont {F.}~\bibnamefont {Bencivenga}}, \bibinfo {author} {\bibfnamefont {V.}~\bibnamefont {Unikandanunni}}, \bibinfo {author} {\bibfnamefont {F.}~\bibnamefont {Capotondi}}, \bibinfo {author} {\bibfnamefont {G.}~\bibnamefont {Chen}}, \bibinfo {author} {\bibfnamefont {Z.}~\bibnamefont {Ding}}, \bibinfo {author} {\bibfnamefont {R.~A.}\ \bibnamefont {Duncan}}, \bibinfo {author} {\bibfnamefont {L.}~\bibnamefont {Foglia}}, \bibinfo {author} {\bibfnamefont {M.~G.}\ \bibnamefont {Izzo}}, \bibinfo {author} {\bibfnamefont {C.}~\bibnamefont {Masciovecchio}}, \bibinfo {author} {\bibfnamefont {A.}~\bibnamefont {Martinelli}}, \bibinfo {author} {\bibfnamefont {G.}~\bibnamefont {Monaco}}, \bibinfo {author} {\bibfnamefont {E.}~\bibnamefont {Pedersoli}}, \bibinfo {author} {\bibfnamefont {S.}~\bibnamefont {Bonetti}},\ and\ \bibinfo {author}
  {\bibfnamefont {K.~A.}\ \bibnamefont {Nelson}},\ }\bibfield  {title} {\bibinfo {title} {Generation and detection of 50 {GHz} surface acoustic waves by extreme ultraviolet pulses},\ }\href {https://doi.org/10.1063/5.0060575} {\bibfield  {journal} {\bibinfo  {journal} {Appl. Phys. Lett.}\ }\textbf {\bibinfo {volume} {119}},\ \bibinfo {pages} {044102} (\bibinfo {year} {2021})}\BibitemShut {NoStop}%
\bibitem [{\citenamefont {Rostami}\ and\ \citenamefont {Asgari}(2015)}]{Rostami2015-zl}%
  \BibitemOpen
  \bibfield  {author} {\bibinfo {author} {\bibfnamefont {H.}~\bibnamefont {Rostami}}\ and\ \bibinfo {author} {\bibfnamefont {R.}~\bibnamefont {Asgari}},\ }\bibfield  {title} {\bibinfo {title} {{Charge compressibility and quantum magnetic phase transition in ${\mathrm{MoS}}_{2}$}},\ }\href {https://doi.org/10.1103/physrevb.91.235301} {\bibfield  {journal} {\bibinfo  {journal} {Phys. Rev. B.}\ }\textbf {\bibinfo {volume} {91}},\ \bibinfo {pages} {235301} (\bibinfo {year} {2015})}\BibitemShut {NoStop}%
\bibitem [{\citenamefont {Kaasbjerg}\ \emph {et~al.}(2019)\citenamefont {Kaasbjerg}, \citenamefont {Low},\ and\ \citenamefont {Jauho}}]{Kaasbjerg2019-ig}%
  \BibitemOpen
  \bibfield  {author} {\bibinfo {author} {\bibfnamefont {K.}~\bibnamefont {Kaasbjerg}}, \bibinfo {author} {\bibfnamefont {T.}~\bibnamefont {Low}},\ and\ \bibinfo {author} {\bibfnamefont {A.-P.}\ \bibnamefont {Jauho}},\ }\bibfield  {title} {\bibinfo {title} {{Electron and hole transport in disordered monolayer ${\mathrm{MoS}}_{2}$: Atomic vacancy induced short-range and Coulomb disorder scattering}},\ }\href {https://doi.org/10.1103/physrevb.100.115409} {\bibfield  {journal} {\bibinfo  {journal} {Phys. Rev. B.}\ }\textbf {\bibinfo {volume} {100}},\ \bibinfo {pages} {115409} (\bibinfo {year} {2019})}\BibitemShut {NoStop}%
\bibitem [{\citenamefont {Sun}\ \emph {et~al.}(2020)\citenamefont {Sun}, \citenamefont {Fogler}, \citenamefont {Basov},\ and\ \citenamefont {Millis}}]{Sun2020-tw}%
  \BibitemOpen
  \bibfield  {author} {\bibinfo {author} {\bibfnamefont {Z.}~\bibnamefont {Sun}}, \bibinfo {author} {\bibfnamefont {M.~M.}\ \bibnamefont {Fogler}}, \bibinfo {author} {\bibfnamefont {D.~N.}\ \bibnamefont {Basov}},\ and\ \bibinfo {author} {\bibfnamefont {A.~J.}\ \bibnamefont {Millis}},\ }\bibfield  {title} {\bibinfo {title} {{Collective modes and terahertz near-field response of superconductors}},\ }\href {https://doi.org/10.1103/physrevresearch.2.023413} {\bibfield  {journal} {\bibinfo  {journal} {Phys. Rev. Res.}\ }\textbf {\bibinfo {volume} {2}},\ \bibinfo {pages} {023413} (\bibinfo {year} {2020})}\BibitemShut {NoStop}%
\bibitem [{\citenamefont {Bénédic}\ \emph {et~al.}(2004)\citenamefont {Bénédic}, \citenamefont {Assouar}, \citenamefont {Mohasseb}, \citenamefont {Elmazria}, \citenamefont {Alnot},\ and\ \citenamefont {Gicquel}}]{Benedic2004-rg}%
  \BibitemOpen
  \bibfield  {author} {\bibinfo {author} {\bibfnamefont {F.}~\bibnamefont {Bénédic}}, \bibinfo {author} {\bibfnamefont {M.~B.}\ \bibnamefont {Assouar}}, \bibinfo {author} {\bibfnamefont {F.}~\bibnamefont {Mohasseb}}, \bibinfo {author} {\bibfnamefont {O.}~\bibnamefont {Elmazria}}, \bibinfo {author} {\bibfnamefont {P.}~\bibnamefont {Alnot}},\ and\ \bibinfo {author} {\bibfnamefont {A.}~\bibnamefont {Gicquel}},\ }\bibfield  {title} {\bibinfo {title} {Surface acoustic wave devices based on nanocrystalline diamond and aluminium nitride},\ }\href {https://doi.org/10.1016/j.diamond.2003.10.020} {\bibfield  {journal} {\bibinfo  {journal} {Diam. Relat. Mater.}\ }\textbf {\bibinfo {volume} {13}},\ \bibinfo {pages} {347} (\bibinfo {year} {2004})}\BibitemShut {NoStop}%
\bibitem [{\citenamefont {Rodriguez-Madrid}\ \emph {et~al.}(2012)\citenamefont {Rodriguez-Madrid}, \citenamefont {Iriarte}, \citenamefont {Pedros}, \citenamefont {Williams}, \citenamefont {Brink},\ and\ \citenamefont {Calle}}]{Rodriguez-Madrid2012-dj}%
  \BibitemOpen
  \bibfield  {author} {\bibinfo {author} {\bibfnamefont {J.~G.}\ \bibnamefont {Rodriguez-Madrid}}, \bibinfo {author} {\bibfnamefont {G.~F.}\ \bibnamefont {Iriarte}}, \bibinfo {author} {\bibfnamefont {J.}~\bibnamefont {Pedros}}, \bibinfo {author} {\bibfnamefont {O.~A.}\ \bibnamefont {Williams}}, \bibinfo {author} {\bibfnamefont {D.}~\bibnamefont {Brink}},\ and\ \bibinfo {author} {\bibfnamefont {F.}~\bibnamefont {Calle}},\ }\bibfield  {title} {\bibinfo {title} {{Super-High-Frequency SAW Resonators on AlN/Diamond}},\ }\href {https://doi.org/10.1109/led.2012.2183851} {\bibfield  {journal} {\bibinfo  {journal} {IEEE Electron Device Lett.}\ }\textbf {\bibinfo {volume} {33}},\ \bibinfo {pages} {495} (\bibinfo {year} {2012})}\BibitemShut {NoStop}%
\bibitem [{\citenamefont {Lillie}\ \emph {et~al.}(2019)\citenamefont {Lillie}, \citenamefont {Dontschuk}, \citenamefont {Broadway}, \citenamefont {Creedon}, \citenamefont {Hollenberg},\ and\ \citenamefont {Tetienne}}]{Lillie2019-tv}%
  \BibitemOpen
  \bibfield  {author} {\bibinfo {author} {\bibfnamefont {S.~E.}\ \bibnamefont {Lillie}}, \bibinfo {author} {\bibfnamefont {N.}~\bibnamefont {Dontschuk}}, \bibinfo {author} {\bibfnamefont {D.~A.}\ \bibnamefont {Broadway}}, \bibinfo {author} {\bibfnamefont {D.~L.}\ \bibnamefont {Creedon}}, \bibinfo {author} {\bibfnamefont {L.~C.~L.}\ \bibnamefont {Hollenberg}},\ and\ \bibinfo {author} {\bibfnamefont {J.-P.}\ \bibnamefont {Tetienne}},\ }\bibfield  {title} {\bibinfo {title} {{Imaging Graphene Field-Effect Transistors on Diamond Using Nitrogen-Vacancy Microscopy}},\ }\href {https://doi.org/10.1103/PhysRevApplied.12.024018} {\bibfield  {journal} {\bibinfo  {journal} {Phys. Rev. Appl.}\ }\textbf {\bibinfo {volume} {12}},\ \bibinfo {pages} {024018} (\bibinfo {year} {2019})}\BibitemShut {NoStop}%
\bibitem [{\citenamefont {Zhou}\ \emph {et~al.}(2020)\citenamefont {Zhou}, \citenamefont {Jerger}, \citenamefont {Lee}, \citenamefont {Fukami}, \citenamefont {Mujid}, \citenamefont {Park},\ and\ \citenamefont {Awschalom}}]{Zhou2020-kb}%
  \BibitemOpen
  \bibfield  {author} {\bibinfo {author} {\bibfnamefont {B.~B.}\ \bibnamefont {Zhou}}, \bibinfo {author} {\bibfnamefont {P.~C.}\ \bibnamefont {Jerger}}, \bibinfo {author} {\bibfnamefont {K.-H.}\ \bibnamefont {Lee}}, \bibinfo {author} {\bibfnamefont {M.}~\bibnamefont {Fukami}}, \bibinfo {author} {\bibfnamefont {F.}~\bibnamefont {Mujid}}, \bibinfo {author} {\bibfnamefont {J.}~\bibnamefont {Park}},\ and\ \bibinfo {author} {\bibfnamefont {D.~D.}\ \bibnamefont {Awschalom}},\ }\bibfield  {title} {\bibinfo {title} {{Spatiotemporal Mapping of a Photocurrent Vortex in Monolayer MoS$_2$ Using Diamond Quantum Sensors}},\ }\href {https://doi.org/10.1103/physrevx.10.011003} {\bibfield  {journal} {\bibinfo  {journal} {Phys. Rev. X.}\ }\textbf {\bibinfo {volume} {10}},\ \bibinfo {pages} {011003} (\bibinfo {year} {2020})}\BibitemShut {NoStop}%
\bibitem [{\citenamefont {Lin}\ \emph {et~al.}(2021)\citenamefont {Lin}, \citenamefont {Ong}, \citenamefont {Bange}, \citenamefont {Faria~Junior}, \citenamefont {Peng}, \citenamefont {Ziegler}, \citenamefont {Zipfel}, \citenamefont {B{\"{a}}uml}, \citenamefont {Paradiso}, \citenamefont {Watanabe}, \citenamefont {Taniguchi}, \citenamefont {Strunk}, \citenamefont {Monserrat}, \citenamefont {Fabian}, \citenamefont {Chernikov}, \citenamefont {Qiu}, \citenamefont {Louie},\ and\ \citenamefont {Lupton}}]{Lin2021-rs}%
  \BibitemOpen
  \bibfield  {author} {\bibinfo {author} {\bibfnamefont {K.-Q.}\ \bibnamefont {Lin}}, \bibinfo {author} {\bibfnamefont {C.~S.}\ \bibnamefont {Ong}}, \bibinfo {author} {\bibfnamefont {S.}~\bibnamefont {Bange}}, \bibinfo {author} {\bibfnamefont {P.~E.}\ \bibnamefont {Faria~Junior}}, \bibinfo {author} {\bibfnamefont {B.}~\bibnamefont {Peng}}, \bibinfo {author} {\bibfnamefont {J.~D.}\ \bibnamefont {Ziegler}}, \bibinfo {author} {\bibfnamefont {J.}~\bibnamefont {Zipfel}}, \bibinfo {author} {\bibfnamefont {C.}~\bibnamefont {B{\"{a}}uml}}, \bibinfo {author} {\bibfnamefont {N.}~\bibnamefont {Paradiso}}, \bibinfo {author} {\bibfnamefont {K.}~\bibnamefont {Watanabe}}, \bibinfo {author} {\bibfnamefont {T.}~\bibnamefont {Taniguchi}}, \bibinfo {author} {\bibfnamefont {C.}~\bibnamefont {Strunk}}, \bibinfo {author} {\bibfnamefont {B.}~\bibnamefont {Monserrat}}, \bibinfo {author} {\bibfnamefont {J.}~\bibnamefont {Fabian}}, \bibinfo {author} {\bibfnamefont {A.}~\bibnamefont {Chernikov}}, \bibinfo {author} {\bibfnamefont
  {D.~Y.}\ \bibnamefont {Qiu}}, \bibinfo {author} {\bibfnamefont {S.~G.}\ \bibnamefont {Louie}},\ and\ \bibinfo {author} {\bibfnamefont {J.~M.}\ \bibnamefont {Lupton}},\ }\bibfield  {title} {\bibinfo {title} {{Narrow-band high-lying excitons with negative-mass electrons in monolayer {WSe$_2$}}},\ }\href {https://doi.org/10.1038/s41467-021-25499-2} {\bibfield  {journal} {\bibinfo  {journal} {Nat. Commun.}\ }\textbf {\bibinfo {volume} {12}},\ \bibinfo {pages} {5500} (\bibinfo {year} {2021})}\BibitemShut {NoStop}%
\bibitem [{\citenamefont {Zeng}\ \emph {et~al.}(2023)\citenamefont {Zeng}, \citenamefont {Xing}, \citenamefont {Wu}, \citenamefont {Li}, \citenamefont {Huang},\ and\ \citenamefont {Liu}}]{Zeng2023-wx}%
  \BibitemOpen
  \bibfield  {author} {\bibinfo {author} {\bibfnamefont {S.-S.}\ \bibnamefont {Zeng}}, \bibinfo {author} {\bibfnamefont {Y.-Q.}\ \bibnamefont {Xing}}, \bibinfo {author} {\bibfnamefont {Z.}~\bibnamefont {Wu}}, \bibinfo {author} {\bibfnamefont {B.-J.}\ \bibnamefont {Li}}, \bibinfo {author} {\bibfnamefont {P.}~\bibnamefont {Huang}},\ and\ \bibinfo {author} {\bibfnamefont {L.}~\bibnamefont {Liu}},\ }\bibfield  {title} {\bibinfo {title} {{Enhanced Energy Transfer between Nitrogen-Vacancy Centers and 2D {MoS}$_{2}$ Films Accurately Fabricated by Atomic Layer Deposition}},\ }\href {https://doi.org/10.1002/adom.202203105} {\bibfield  {journal} {\bibinfo  {journal} {Adv. Opt. Mater.}\ }\textbf {\bibinfo {volume} {11}},\ \bibinfo {pages} {2203105} (\bibinfo {year} {2023})}\BibitemShut {NoStop}%
\bibitem [{\citenamefont {Ko\v{c}\'{\i}}\ \emph {et~al.}(2023)\citenamefont {Ko\v{c}\'{\i}}, \citenamefont {Izs\'{a}k}, \citenamefont {Vanko}, \citenamefont {Sojkov\'{a}}, \citenamefont {Hrd\'{a}}, \citenamefont {Szab\'{o}}, \citenamefont {Hus\'{a}k}, \citenamefont {V\'{e}gs{\"{o}}}, \citenamefont {Varga},\ and\ \citenamefont {Kromka}}]{Koci2023-gb}%
  \BibitemOpen
  \bibfield  {author} {\bibinfo {author} {\bibfnamefont {M.}~\bibnamefont {Ko\v{c}\'{\i}}}, \bibinfo {author} {\bibfnamefont {T.}~\bibnamefont {Izs\'{a}k}}, \bibinfo {author} {\bibfnamefont {G.}~\bibnamefont {Vanko}}, \bibinfo {author} {\bibfnamefont {M.}~\bibnamefont {Sojkov\'{a}}}, \bibinfo {author} {\bibfnamefont {J.}~\bibnamefont {Hrd\'{a}}}, \bibinfo {author} {\bibfnamefont {O.}~\bibnamefont {Szab\'{o}}}, \bibinfo {author} {\bibfnamefont {M.}~\bibnamefont {Hus\'{a}k}}, \bibinfo {author} {\bibfnamefont {K.}~\bibnamefont {V\'{e}gs{\"{o}}}}, \bibinfo {author} {\bibfnamefont {M.}~\bibnamefont {Varga}},\ and\ \bibinfo {author} {\bibfnamefont {A.}~\bibnamefont {Kromka}},\ }\bibfield  {title} {\bibinfo {title} {{Improved Gas Sensing Capabilities of {MoS$_2$}/Diamond Heterostructures at Room Temperature}},\ }\href {https://doi.org/10.1021/acsami.3c04438} {\bibfield  {journal} {\bibinfo  {journal} {ACS Appl. Mater. Interfaces}\ }\textbf {\bibinfo {volume} {15}},\ \bibinfo {pages} {34206} (\bibinfo {year}
  {2023})}\BibitemShut {NoStop}%
\bibitem [{\citenamefont {Fescenko}\ \emph {et~al.}(2025)\citenamefont {Fescenko}, \citenamefont {Kumar}, \citenamefont {Gas-Osoth}, \citenamefont {Wang}, \citenamefont {Lamichhane}, \citenamefont {Li}, \citenamefont {Erickson}, \citenamefont {Raghavan}, \citenamefont {Delord}, \citenamefont {Cress}, \citenamefont {Proscia}, \citenamefont {LaGasse}, \citenamefont {Liou}, \citenamefont {Hong}, \citenamefont {Fonseca}, \citenamefont {An}, \citenamefont {Meriles},\ and\ \citenamefont {Laraoui}}]{Fescenko2025-sl}%
  \BibitemOpen
  \bibfield  {author} {\bibinfo {author} {\bibfnamefont {I.}~\bibnamefont {Fescenko}}, \bibinfo {author} {\bibfnamefont {R.}~\bibnamefont {Kumar}}, \bibinfo {author} {\bibfnamefont {T.}~\bibnamefont {Gas-Osoth}}, \bibinfo {author} {\bibfnamefont {Y.}~\bibnamefont {Wang}}, \bibinfo {author} {\bibfnamefont {S.}~\bibnamefont {Lamichhane}}, \bibinfo {author} {\bibfnamefont {T.}~\bibnamefont {Li}}, \bibinfo {author} {\bibfnamefont {A.}~\bibnamefont {Erickson}}, \bibinfo {author} {\bibfnamefont {N.}~\bibnamefont {Raghavan}}, \bibinfo {author} {\bibfnamefont {T.}~\bibnamefont {Delord}}, \bibinfo {author} {\bibfnamefont {C.~D.}\ \bibnamefont {Cress}}, \bibinfo {author} {\bibfnamefont {N.}~\bibnamefont {Proscia}}, \bibinfo {author} {\bibfnamefont {S.~W.}\ \bibnamefont {LaGasse}}, \bibinfo {author} {\bibfnamefont {S.-H.}\ \bibnamefont {Liou}}, \bibinfo {author} {\bibfnamefont {X.}~\bibnamefont {Hong}}, \bibinfo {author} {\bibfnamefont {J.~J.}\ \bibnamefont {Fonseca}}, \bibinfo {author} {\bibfnamefont {T.}~\bibnamefont
  {An}}, \bibinfo {author} {\bibfnamefont {C.~A.}\ \bibnamefont {Meriles}},\ and\ \bibinfo {author} {\bibfnamefont {A.}~\bibnamefont {Laraoui}},\ }\bibfield  {title} {\bibinfo {title} {{Nitrogen-Vacancy Magnetometry of Edge Magnetism in {WS}$_{2}$ Flakes}},\ }\href {https://doi.org/10.1002/adfm.202512391} {\bibfield  {journal} {\bibinfo  {journal} {Adv. Funct. Mater.}\ }\textbf {\bibinfo {volume} {35}},\ \bibinfo {pages} {e12391} (\bibinfo {year} {2025})}\BibitemShut {NoStop}%
\bibitem [{\citenamefont {Yoshioka}\ \emph {et~al.}(2022)\citenamefont {Yoshioka}, \citenamefont {Wakamura}, \citenamefont {Hashisaka}, \citenamefont {Watanabe}, \citenamefont {Taniguchi},\ and\ \citenamefont {Kumada}}]{Yoshioka2022-yr}%
  \BibitemOpen
  \bibfield  {author} {\bibinfo {author} {\bibfnamefont {K.}~\bibnamefont {Yoshioka}}, \bibinfo {author} {\bibfnamefont {T.}~\bibnamefont {Wakamura}}, \bibinfo {author} {\bibfnamefont {M.}~\bibnamefont {Hashisaka}}, \bibinfo {author} {\bibfnamefont {K.}~\bibnamefont {Watanabe}}, \bibinfo {author} {\bibfnamefont {T.}~\bibnamefont {Taniguchi}},\ and\ \bibinfo {author} {\bibfnamefont {N.}~\bibnamefont {Kumada}},\ }\bibfield  {title} {\bibinfo {title} {Ultrafast intrinsic optical-to-electrical conversion dynamics in a graphene photodetector},\ }\href {https://doi.org/10.1038/s41566-022-01058-z} {\bibfield  {journal} {\bibinfo  {journal} {Nat. Photonics}\ }\textbf {\bibinfo {volume} {16}},\ \bibinfo {pages} {718} (\bibinfo {year} {2022})}\BibitemShut {NoStop}%
\bibitem [{\citenamefont {Rees}\ \emph {et~al.}(2020)\citenamefont {Rees}, \citenamefont {Manna}, \citenamefont {Lu}, \citenamefont {Morimoto}, \citenamefont {Borrmann}, \citenamefont {Felser}, \citenamefont {Moore}, \citenamefont {Torchinsky},\ and\ \citenamefont {Orenstein}}]{Rees2020-ws}%
  \BibitemOpen
  \bibfield  {author} {\bibinfo {author} {\bibfnamefont {D.}~\bibnamefont {Rees}}, \bibinfo {author} {\bibfnamefont {K.}~\bibnamefont {Manna}}, \bibinfo {author} {\bibfnamefont {B.}~\bibnamefont {Lu}}, \bibinfo {author} {\bibfnamefont {T.}~\bibnamefont {Morimoto}}, \bibinfo {author} {\bibfnamefont {H.}~\bibnamefont {Borrmann}}, \bibinfo {author} {\bibfnamefont {C.}~\bibnamefont {Felser}}, \bibinfo {author} {\bibfnamefont {J.~E.}\ \bibnamefont {Moore}}, \bibinfo {author} {\bibfnamefont {D.~H.}\ \bibnamefont {Torchinsky}},\ and\ \bibinfo {author} {\bibfnamefont {J.}~\bibnamefont {Orenstein}},\ }\bibfield  {title} {\bibinfo {title} {{Helicity-dependent photocurrents in the chiral Weyl semimetal {RhSi}}},\ }\href {https://doi.org/10.1126/sciadv.aba0509} {\bibfield  {journal} {\bibinfo  {journal} {Sci. Adv.}\ }\textbf {\bibinfo {volume} {6}},\ \bibinfo {pages} {eaba0509} (\bibinfo {year} {2020})}\BibitemShut {NoStop}%
\bibitem [{\citenamefont {Wang}\ \emph {et~al.}(2022)\citenamefont {Wang}, \citenamefont {Ota}, \citenamefont {Edlbauer}, \citenamefont {Jadot}, \citenamefont {Mortemousque}, \citenamefont {Richard}, \citenamefont {Okazaki}, \citenamefont {Nakamura}, \citenamefont {Ludwig}, \citenamefont {Wieck}, \citenamefont {Urdampilleta}, \citenamefont {Meunier}, \citenamefont {Kodera}, \citenamefont {Kaneko}, \citenamefont {Takada},\ and\ \citenamefont {B{\"{a}}uerle}}]{Wang2022-es}%
  \BibitemOpen
  \bibfield  {author} {\bibinfo {author} {\bibfnamefont {J.}~\bibnamefont {Wang}}, \bibinfo {author} {\bibfnamefont {S.}~\bibnamefont {Ota}}, \bibinfo {author} {\bibfnamefont {H.}~\bibnamefont {Edlbauer}}, \bibinfo {author} {\bibfnamefont {B.}~\bibnamefont {Jadot}}, \bibinfo {author} {\bibfnamefont {P.-A.}\ \bibnamefont {Mortemousque}}, \bibinfo {author} {\bibfnamefont {A.}~\bibnamefont {Richard}}, \bibinfo {author} {\bibfnamefont {Y.}~\bibnamefont {Okazaki}}, \bibinfo {author} {\bibfnamefont {S.}~\bibnamefont {Nakamura}}, \bibinfo {author} {\bibfnamefont {A.}~\bibnamefont {Ludwig}}, \bibinfo {author} {\bibfnamefont {A.~D.}\ \bibnamefont {Wieck}}, \bibinfo {author} {\bibfnamefont {M.}~\bibnamefont {Urdampilleta}}, \bibinfo {author} {\bibfnamefont {T.}~\bibnamefont {Meunier}}, \bibinfo {author} {\bibfnamefont {T.}~\bibnamefont {Kodera}}, \bibinfo {author} {\bibfnamefont {N.-H.}\ \bibnamefont {Kaneko}}, \bibinfo {author} {\bibfnamefont {S.}~\bibnamefont {Takada}},\ and\ \bibinfo {author} {\bibfnamefont
  {C.}~\bibnamefont {B{\"{a}}uerle}},\ }\bibfield  {title} {\bibinfo {title} {{Generation of a Single-Cycle Acoustic Pulse: A Scalable Solution for Transport in Single-Electron Circuits}},\ }\href {https://doi.org/10.1103/physrevx.12.031035} {\bibfield  {journal} {\bibinfo  {journal} {Phys. Rev. X.}\ }\textbf {\bibinfo {volume} {12}},\ \bibinfo {pages} {031035} (\bibinfo {year} {2022})}\BibitemShut {NoStop}%
\bibitem [{\citenamefont {Fujiwara}\ \emph {et~al.}(2025)\citenamefont {Fujiwara}, \citenamefont {Ota}, \citenamefont {Kodera}, \citenamefont {Okazaki}, \citenamefont {Kaneko}, \citenamefont {Jiang}, \citenamefont {Niimi},\ and\ \citenamefont {Takada}}]{Fujiwara2025-rm}%
  \BibitemOpen
  \bibfield  {author} {\bibinfo {author} {\bibfnamefont {K.}~\bibnamefont {Fujiwara}}, \bibinfo {author} {\bibfnamefont {S.}~\bibnamefont {Ota}}, \bibinfo {author} {\bibfnamefont {T.}~\bibnamefont {Kodera}}, \bibinfo {author} {\bibfnamefont {Y.}~\bibnamefont {Okazaki}}, \bibinfo {author} {\bibfnamefont {N.-H.}\ \bibnamefont {Kaneko}}, \bibinfo {author} {\bibfnamefont {N.}~\bibnamefont {Jiang}}, \bibinfo {author} {\bibfnamefont {Y.}~\bibnamefont {Niimi}},\ and\ \bibinfo {author} {\bibfnamefont {S.}~\bibnamefont {Takada}},\ }\bibfield  {title} {\bibinfo {title} {Generation of a single-cycle surface acoustic wave pulse on {LiNbO$_3$} for application to thin-film materials},\ }\href {https://doi.org/10.1063/5.0270260} {\bibfield  {journal} {\bibinfo  {journal} {Appl. Phys. Lett.}\ }\textbf {\bibinfo {volume} {127}},\ \bibinfo {pages} {022406} (\bibinfo {year} {2025})}\BibitemShut {NoStop}%
\end{thebibliography}%

\clearpage
\onecolumngrid\begin{center}\textbf{End Matter}\end{center}
\twocolumngrid
\textit{The details of continuum model}.---
Here, we provide the details of the continuum model introduced in Eq.~\eqref{Eq:continuum}.
Each component of the continuum model is given by
\begin{subequations}
\begin{align}
    {H}_{\tau,l}(\hat{\bm{k}})
    &= -\frac{\hbar^{2}}{2m^{*}}(\hat{\bm{k}}-\bm{\kappa}_{\tau,l})^{2} - l\frac{V_{z}}{2}{-\mu}, \\
    \Delta_{l}(\hat{\bm{r}})
    &= V\sum_{j=1,2,3}e^{i(\bm{g}_{j}\cdot\hat{\bm{r}}+l\psi)} + {\rm h.c.}, \\
    T_{\tau}(\hat{\bm{r}})
    &= w(1+e^{+i\tau\bm{g}_{2}\cdot\hat{\bm{r}}}+e^{-i\tau\bm{g}_{3}\cdot\hat{\bm{r}}}),
\end{align}
\end{subequations}
with the effective mass $m^\ast=0.43\,m_{e}$, the momentum shifts $\bm{\kappa}_{\tau,l} = [4\pi\tau/3L_{M}](\sqrt{3}/2,l/2)$, layer potential difference $V_z$, the chemical potential $\mu$, the intralayer coupling strength $V$, the phase offset $\psi$ and the interlayer coupling strength $w$.
Here, the moir\'e lattice constant and the reciprocal lattice vectors in the MBZ are given by, $L_{M} = a_{0}/[2\sin\frac{\theta}{2}]$ and $\bm{g}_{j} = \frac{4\pi}{\sqrt{3}L_{M}}\qty(\cos\frac{2(j-1)\pi}{3},\;\sin\frac{2(j-1)\pi}{3})$.
The three vectors $\qty{\bm{g}_j}$ are the shortest reciprocal vectors at 120$^{\circ}$ to each other; the MBZ is spanned by $\bm{g}_1$ and $\bm{g}_2$, while $\bm{g}_3$ satisfies $\bm{g}_1 + \bm{g}_2 = -\bm{g}_3$.
Throughout this Letter, we set the parameters:
$\qty(\theta, \psi, w, V, V_{z},\mu) = \qty(5^\circ, 128^\circ, 18\;{\rm meV}, 9\;{\rm meV}, 43.75\;{\rm meV}, -13\;{\rm meV})$,
as obtained from experimental configuration and density functional theory calculations~\cite{Devakul2021-hs,Guo2025-kt,Zhu2025-ba}
and the lattice constant of each monolayer $a_{0} = 3.317 \; \text{\AA}$.

\textit{Microscopic derivation of strain-induced couplings}.---
Here, we discuss the microscopic origin and relative importance of the strain couplings induced by SAWs.
In Sec.~S2 of the Supplemental Material, we analyze these couplings using a Slater-Koster tight-binding model for strained transition-metal dichalcogenides~\cite{Rostami2015-hh} and project them onto the relevant low-energy valence band of tWSe$_2$.
The projection identifies the pseudo-gauge field as the leading valley-odd coupling and the deformation potential as the leading valley-even scalar coupling.
The projected pseudo-gauge field coincides with $\bm{A}_{s}$ employed in the main text.
By contrast, the deformation potential couples to charge rather than to the valley degree of freedom and generates a distinct rectified dc response, which we refer to as the SAEE.
The SAEE is analyzed separately in Sec.~S7.1 of the Supplemental Material~\cite{Supplemental}.
Other strain-induced couplings are classified and estimated in Table~S1 of the Supplemental Material and are subleading for the Rayleigh-wave setup considered here.
Thus, the deformation-potential-driven SAEE constitutes the principal competing contribution to the SAGE.
When a two-dimensional system embedded in three dimensions is considered, the scalar deformation-potential coupling is strongly screened~\cite{Rostami2015-zl,Kaasbjerg2019-ig,Sun2020-tw}, favoring the pseudo-gauge-field contribution.
In a gated geometry, however, the relative magnitude of the SAGE and SAEE becomes device dependent, particularly through the gate distance.
Increasing the gate distance restores the long-range character of the Coulomb interaction and thereby suppresses the SAEE relative to the SAGE.

At the level of the injection response, the SAEE has the same formal structure as the SAGE, with the pseudo-gauge-field vertices replaced by deformation-potential vertices.
Consequently, in a nodal superconducting state, the corresponding coupling-dependent quantum-geometric factor can produce a characteristic low-frequency enhancement even when the SAEE is comparable to or larger than the SAGE.
The low-frequency enhancement and the emergence of nematicity below $T_{\rm c}$ therefore remain useful diagnostics of the superconducting gap structure, whereas attributing the measured dc current specifically to the SAGE requires separating the relevant strain-coupling channels.
One possible strategy is to use a transverse acoustic mode, such as a Love wave, which eliminates the leading linear deformation-potential coupling and thereby isolates the SAGE, although longitudinal Rayleigh waves are generally more accessible in experiments.
Further details are provided in Sec.~S7.1 of the Supplemental Material~\cite{Supplemental}.

\begin{figure*}[tbp]
    \centering
    \includegraphics[width=\linewidth,keepaspectratio]{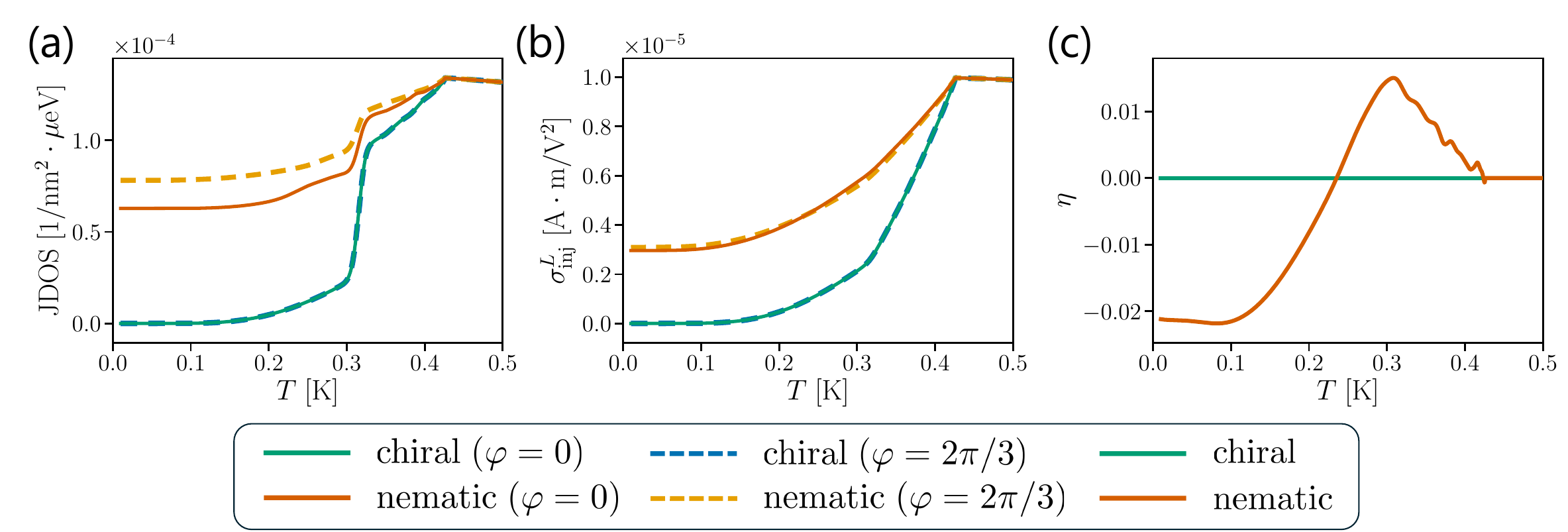}
    \caption{Temperature dependence of the JDOS, acoustogalvanic conductivity $\sigma^L_{\mathrm{inj}}$, and nematicity $\eta$. Green and orange curves correspond to chiral and nematic states, respectively. In (a) and (b), solid and dashed lines indicate the azimuthal angles $\varphi=0$ and $2\pi/3$, respectively; the results for the chiral states coincide by symmetry. Parameters are fixed at $\Omega/2\pi = 25\,\mathrm{GHz}$.}
    \label{fig:plot_d_temp}
\end{figure*}
\textit{Temperature dependence of SAGE for chiral/nematic $d$-wave states}.---
Here, we present the temperature dependence of SAGE.
Figure~\ref{fig:plot_d_temp} shows the temperature dependence of the JDOS, acoustogalvanic conductivity, and nematicity for chiral and nematic $d$-wave states.
We first consider the JDOS at a fixed frequency $\hbar\Omega = 100\,\mu\mathrm{eV}$, as shown in Fig.~\ref{fig:plot_d_temp}(a).
This fixed frequency corresponds to SAWs with $\Omega/2\pi=25\,\mathrm{GHz}$ and wavelength $\lambda_{\rm SAW}\simeq160\,\mathrm{nm}$, which is a few times longer than the coherence length of tWSe$_2$, $\xi_0 \sim 57 \,\mathrm{nm}$~\cite{Guo2025-kt}.
The JDOS remains finite in both the chiral and nematic states as long as $\hbar\Omega$ exceeds the characteristic excitation threshold $2\Delta(T)$, which increases upon cooling.
At lower temperatures, however, the two states behave markedly differently.
In the chiral state, the JDOS rapidly diminishes because of its fully gapped spectrum.
By contrast, in the nematic state, the nodal and anisotropic gap structure sustains a finite JDOS, allowing low-energy quasiparticles to remain resonant.

Guided by this JDOS analysis, we now turn to the temperature dependence of SAGE shown in Fig.~\ref{fig:plot_d_temp}(b).
In the chiral state, the onset of the acoustogalvanic conductivity follows JDOS: a sizable response appears only in the temperature range where the resonance condition is satisfied.
By contrast, in the nematic state, a finite response persists throughout the superconducting phase because nodal quasiparticles remain available for resonant transitions.
Moreover, Fig.~\ref{fig:plot_d_temp}(c) shows that only the nematic state exhibits a finite nematicity, signaling spontaneous rotational symmetry breaking.
These distinct temperature dependences of SAGE and nematicity provide a sensitive probe of low-energy quasiparticles and offer a possible route to experimentally distinguish nematic and chiral superconducting states.

\begin{figure}[tbp]
    \centering
    \includegraphics[width=\linewidth,keepaspectratio]{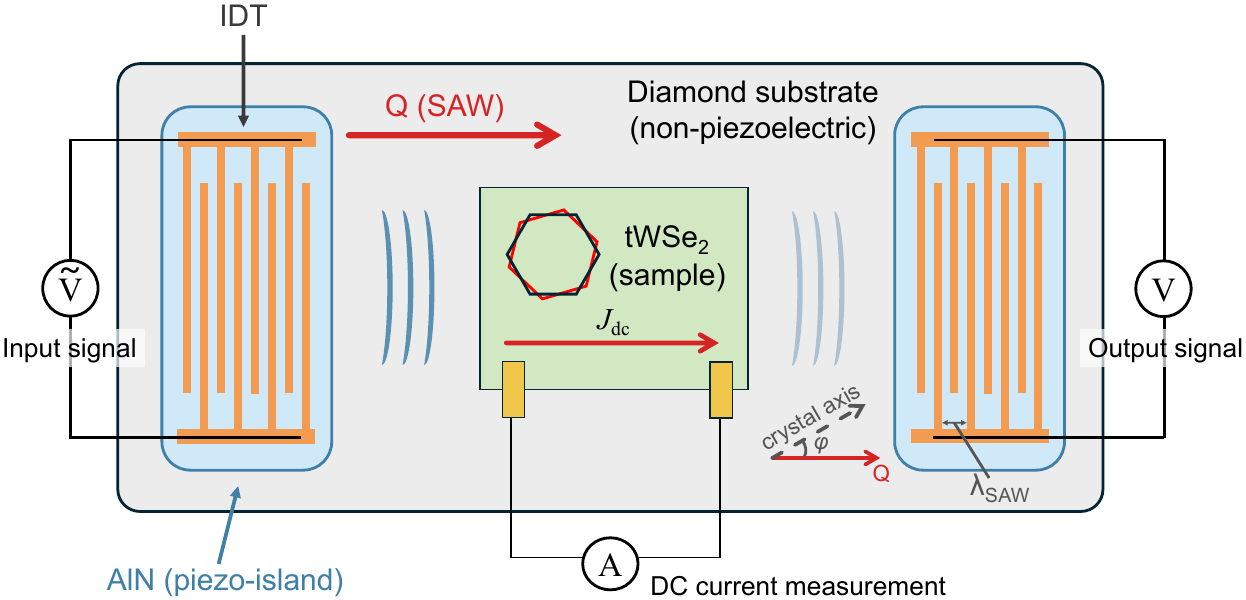}
    \caption{Schematic illustration of the experimental setup for measuring the SAGE (top view). Periodic electric signals applied to the interdigital transducers (IDTs) excite SAWs with wave vector $\bm{Q}$ via piezoelectric coupling. The IDTs are fabricated on AlN piezoelectric islands placed on a non-piezoelectric diamond substrate, thereby suppressing piezoelectric-field-induced responses while retaining strain-induced responses, including the SAGE and the SAEE. The tWSe$_2$ sample is transferred at the center, and the rectified DC current $J_{\rm dc}$ is measured through the sample contacts, while the transmitted wave is monitored by the detection IDT. The propagation direction $\varphi$ is defined relative to the crystalline axis; angle-resolved measurements (e.g., $\varphi=0$ and $2\pi/3$) access the nematicity $\eta$ of the response.}
    \label{fig:setup}
\end{figure}
\textit{Detailed experimental configurations for SAGE}.--- 
We briefly discuss the experimental setup suitable for suppressing piezoelectric-field-induced backgrounds and detecting strain-induced dc responses, as depicted in Fig.~\ref{fig:setup}.
In general, standard elastic substrates such as LiNbO$_3$ not only generate a strain field, which triggers the SAGE, but also a piezoelectric field, known as the dominant source of the acoustoelectric effect~\cite{Fal-ko1993-dz,Miseikis2012-zp,Bandhu2013-qq}
The latter effect can be suppressed by using a non-piezoelectric substrate, such as diamond, with piezoelectric islands and transducers (e.g., AlN) on its surface to generate SAWs~\cite{Benedic2004-rg,Rodriguez-Madrid2012-dj}.
Note that the transfer and integration of vdW materials onto diamond substrates has already been demonstrated~\cite{Lillie2019-tv,Zhou2020-kb,Lin2021-rs,Zeng2023-wx,Koci2023-gb,Fescenko2025-sl}.
In this configuration, piezoelectric-field-driven acoustoelectric currents in the sample are suppressed, whereas strain-induced responses, including both the SAGE and the SAEE, may remain.
Their relative contributions can be controlled through Coulomb screening, gate geometry, and acoustic-wave polarization, as discussed in the main text and Sec.~S7 of the Supplemental Material~\cite{Supplemental}.
The SAW-induced electric current may be detected as a shift of current-voltage characteristic along the current axis, which may in turn lead to an electromotive force e.g., through the intrinsic nonlinear resistance in two-dimensional superconductors. 
A direct measurement  of the SAGE current could also be achieved by a time-resolved current measurement technique~\cite{Yoshioka2022-yr} or by observing the radiation~\cite{Rees2020-ws} in response to the pulsed surface acoustic wave~\cite{Wang2022-es,Fujiwara2025-rm}.

%
%
%
%
%
\clearpage
\ifarXiv
  

\section*{Supplementary material}

## Supplemental Material

### Moiré wave-number representation of the continuum model

In this section, we introduce the moiré wave-number representation of the continuum model. The original continuum model is given by

$$H_\tau = \int d\bm{r} \bm{\Psi}_\tau^\dagger(\bm{r}) \begin{pmatrix} \hat{H}_{\tau,b}(\hat{\bm{k}}) + \Delta_b(\bm{r}) & T_\tau(\bm{r}) \\ T_\tau^\dagger(\bm{r}) & \hat{H}_{\tau,t}(\hat{\bm{k}}) + \Delta_t(\bm{r}) \end{pmatrix} \bm{\Psi}_\tau(\bm{r}), \tag{10}$$

where the components are given by

$$\hat{H}_{\tau,l}(\hat{\bm{k}}) = -\frac{\hbar^2}{2m^*}(\hat{\bm{k}} - \bm{\kappa}_{\tau,l})^2 - l\frac{V_z}{2}, \tag{11}$$

$$\Delta_l(\bm{r}) = V \sum_{j=1,2,3} e^{i(\bm{g}_j \cdot \bm{r} + l\psi)} + \text{h.c.}, \tag{12}$$

$$T_\tau(\bm{r}) = w(1 + e^{+i\tau \bm{g}_2 \cdot \bm{r}} + e^{-i\tau \bm{g}_3 \cdot \bm{r}}). \tag{13}$$

We define the Fourier transformation of the field operator by

$$\psi_l^\dagger(\bm{r}) = \int_{-\infty}^{\infty} \frac{d\bm{k}}{\sqrt{(2\pi)^2}} e^{-i\bm{k} \cdot \bm{r}} c_l^\dagger(\bm{k}), \tag{14}$$

and obtain

$$\begin{aligned} H_m^\tau &= \int d\bm{r} \psi_l^\dagger(\bm{r}) \hat{H}_{\tau,l}(-i\bm{\nabla}) \psi_l(\bm{r}) \\ &= \int_{-\infty}^{\infty} \frac{d\bm{k}}{(2\pi)^2} c_l^\dagger(\bm{k}) H_{\tau,l}(\bm{k}) c_l(\bm{k}), \end{aligned} \tag{15}$$

$$\begin{aligned} H_\Delta^\tau &= \int d\bm{r} \psi_l^\dagger(\bm{r}) \Delta_l(\bm{r}) \psi_l(\bm{r}) \\ &= V e^{il\psi} \sum_{j=1,2,3} \int d\bm{r} \psi_l^\dagger(\bm{r}) e^{i\bm{g}_j \cdot \bm{r}} \psi_l(\bm{r}) + \text{h.c.} \\ &= \int_{-\infty}^{\infty} \frac{d\bm{k}}{(2\pi)^2} V \sum_j \left[ e^{+il\psi} c_l^\dagger(\bm{k} + \bm{g}_j) c_l(\bm{k}) + e^{-il\psi} c_l^\dagger(\bm{k}) c_l(\bm{k} + \bm{g}_j) \right], \end{aligned} \tag{16}$$

$$\begin{aligned} H_T^\tau &= \int d\bm{r} \psi_b^\dagger(\bm{r}) T_\tau(\bm{r}) \psi_t(\bm{r}) + \text{h.c.} \\ &= \int d\bm{r} \psi_b^\dagger(\bm{r}) w(1 + e^{-i\tau \bm{g}_2 \cdot \bm{r}} + e^{+i\tau \bm{g}_3 \cdot \bm{r}}) \psi_t(\bm{r}) + \text{h.c.} \\ &= w \int_{-\infty}^{\infty} \frac{d\bm{k}}{(2\pi)^2} \left[ c_b^\dagger(\bm{k}) c_t(\bm{k}) + c_b^\dagger(\bm{k}) c_t(\bm{k} - \tau \bm{g}_2) + c_b^\dagger(\bm{k} - \tau \bm{g}_3) c_t(\bm{k}) \right] + \text{h.c.}, \end{aligned} \tag{17}$$

where $\bm{g}_j = \frac{4\pi}{\sqrt{3}L_M}\left(\cos\frac{2(j-1)\pi}{3}, \sin\frac{2(j-1)\pi}{3}\right)$ represents moiré reciprocal lattice vectors. In this work, we only consider the first-order interlayer coupling, then, we choose the basis as

$$\bm{\Psi}_b(\bm{k}) = (c_b(\bm{k}), c_b(\bm{k} + \bm{g}_1), c_b(\bm{k} - \bm{g}_3)), \tag{18}$$
$$\bm{\Psi}_t(\bm{k}) = (c_t(\bm{k}), c_t(\bm{k} + \bm{g}_1), c_t(\bm{k} - \bm{g}_2)), \tag{19}$$

by keeping the states near the original massive Dirac cones of the bottom and top layers, respectively. Here, $\bm{k}$ represents the wave number in the moiré Brillouin zone (MBZ). In the following discussion, we focus on the valley $\tau = +$. By rewriting $H_m$ as

$$H_m^+ = \int_{-\infty}^{\infty} \frac{d\bm{k}}{(2\pi)^2} c_l^\dagger(\bm{k}) H_l(\bm{k}) c_l(\bm{k}) \tag{20}$$

$$= \sum_{n,m \in \mathbb{Z}} \int_{\text{MBZ}} \frac{d\bm{k}}{(2\pi)^2} c_l^\dagger(\bm{k} + n\bm{g}_1 + m\bm{g}_2) H_l(\bm{k} + n\bm{g}_1 + m\bm{g}_2) c_l(\bm{k} + n\bm{g}_1 + m\bm{g}_2), \tag{21}$$

and keeping the states specified by $\boldsymbol{\Psi}_b(\boldsymbol{k})$ and $\boldsymbol{\Psi}_l(\boldsymbol{k})$, we obtain $H_m^+ = H_m^b + H_m^l$ with

$$H_m^b = \int_{\mathrm{MBZ}} \frac{d\boldsymbol{k}}{(2\pi)^2} \boldsymbol{\Psi}_b^\dagger(\boldsymbol{k}) \begin{pmatrix} H_b(\boldsymbol{k}) & 0 & 0 \\ 0 & H_b(\boldsymbol{k}+\boldsymbol{g}_1) & 0 \\ 0 & 0 & H_b(\boldsymbol{k}-\boldsymbol{g}_3) \end{pmatrix} \boldsymbol{\Psi}_b(\boldsymbol{k}) \equiv \int_{\mathrm{MBZ}} \frac{d\boldsymbol{k}}{(2\pi)^2} \boldsymbol{\Psi}_b^\dagger(\boldsymbol{k}) h_m^b(\boldsymbol{k}) \boldsymbol{\Psi}_b(\boldsymbol{k}), \qquad (22)$$

$$H_m^t = \int_{\mathrm{MBZ}} \frac{d\boldsymbol{k}}{(2\pi)^2} \boldsymbol{\Psi}_t^\dagger(\boldsymbol{k}) \begin{pmatrix} H_t(\boldsymbol{k}) & 0 & 0 \\ 0 & H_t(\boldsymbol{k}+\boldsymbol{g}_1) & 0 \\ 0 & 0 & H_t(\boldsymbol{k}-\boldsymbol{g}_2) \end{pmatrix} \boldsymbol{\Psi}_t(\boldsymbol{k}) \equiv \int_{\mathrm{MBZ}} \frac{d\boldsymbol{k}}{(2\pi)^2} \boldsymbol{\Psi}_b^\dagger(\boldsymbol{k}) h_m^t(\boldsymbol{k}) \boldsymbol{\Psi}_b(\boldsymbol{k}). \qquad (23)$$

Here and hereafter in this Supplemental Material, we represent the integral over the MBZ by $\int_{\mathrm{MBZ}} \frac{d\boldsymbol{k}}{(2\pi)^2}$ to emphasize the difference from $\int_{-\infty}^{\infty} \frac{d\boldsymbol{k}}{(2\pi)^2}$, instead of $\int \frac{d\boldsymbol{k}}{(2\pi)^2}$ adopted in the main text. To rewrite $H_\Delta(j)$ and $H_T(j)$, we introduce the decomposition

$$H_\Delta = \sum_j H_\Delta(j), \qquad (24)$$

$$H_T = \sum_j H_T(j), \qquad (25)$$

and then, each components are give by

$$\begin{aligned} H_\Delta(1) &= \int_{-\infty}^{\infty} \frac{d\boldsymbol{k}}{(2\pi)^2} V e^{+il\psi} c_l^\dagger(\boldsymbol{k}+\boldsymbol{g}_1) c_l(\boldsymbol{k}) + \mathrm{h.c.} \\ &= \int_{\mathrm{MBZ}} \frac{d\boldsymbol{k}}{(2\pi)^2} V e^{+il\psi} c_l^\dagger(\boldsymbol{k}+\boldsymbol{g}_1) c_l(\boldsymbol{k}) + \mathrm{h.c.} + (\text{other } \boldsymbol{k}), \end{aligned} \qquad (26)$$

$$\begin{aligned} H_\Delta(2) &= \int_{-\infty}^{\infty} \frac{d\boldsymbol{k}}{(2\pi)^2} V e^{+il\psi} c_l^\dagger(\boldsymbol{k}+\boldsymbol{g}_2) c_l(\boldsymbol{k}) + \mathrm{h.c.} \\ &= \int_{-\infty}^{\infty} \frac{d\boldsymbol{k}}{(2\pi)^2} V e^{+il\psi} c_l^\dagger(\boldsymbol{k}) c_l(\boldsymbol{k}-\boldsymbol{g}_2) + \mathrm{h.c.} \\ &= \int_{\mathrm{MBZ}} \frac{d\boldsymbol{k}}{(2\pi)^2} V e^{+il\psi} \left[ c_l^\dagger(\boldsymbol{k}) c_l(\boldsymbol{k}-\boldsymbol{g}_2) + c_l^\dagger(\boldsymbol{k}-\boldsymbol{g}_3) c_l(\boldsymbol{k}+\boldsymbol{g}_1) \right] + \mathrm{h.c.} + (\text{other } \boldsymbol{k}), \end{aligned} \qquad (27)$$

$$\begin{aligned} H_\Delta(3) &= \int_{-\infty}^{\infty} \frac{d\boldsymbol{k}}{(2\pi)^2} V e^{+il\psi} c_l^\dagger(\boldsymbol{k}+\boldsymbol{g}_3) c_l(\boldsymbol{k}) + \mathrm{h.c.} \\ &= \int_{-\infty}^{\infty} \frac{d\boldsymbol{k}}{(2\pi)^2} V e^{+il\psi} c_l^\dagger(\boldsymbol{k}) c_l(\boldsymbol{k}-\boldsymbol{g}_3) + \mathrm{h.c.} \\ &= \int_{\mathrm{MBZ}} \frac{d\boldsymbol{k}}{(2\pi)^2} V e^{+il\psi} \left[ c_l^\dagger(\boldsymbol{k}) c_l(\boldsymbol{k}-\boldsymbol{g}_3) + c_l^\dagger(\boldsymbol{k}-\boldsymbol{g}_2) c_l(\boldsymbol{k}+\boldsymbol{g}_1) \right] + \mathrm{h.c.} + (\text{other } \boldsymbol{k}), \end{aligned} \qquad (28)$$

and

$$\begin{aligned} H_T(1) &= \int_{-\infty}^{\infty} \frac{d\boldsymbol{k}}{(2\pi)^2} w c_b^\dagger(\boldsymbol{k}) c_t(\boldsymbol{k}) + \mathrm{h.c.} \\ &= \int_{\mathrm{MBZ}} \frac{d\boldsymbol{k}}{(2\pi)^2} w \left[ c_b^\dagger(\boldsymbol{k}) c_t(\boldsymbol{k}) + c_b^\dagger(\boldsymbol{k}+\boldsymbol{g}_1) c_t(\boldsymbol{k}+\boldsymbol{g}_1) \right. \\ &\quad \left. + c_b^\dagger(\boldsymbol{k}-\boldsymbol{g}_3) c_t(\boldsymbol{k}-\boldsymbol{g}_3) + c_b^\dagger(\boldsymbol{k}-\boldsymbol{g}_2) c_t(\boldsymbol{k}-\boldsymbol{g}_2) \right] + \mathrm{h.c.} + (\text{other } \boldsymbol{k}), \end{aligned} \qquad (29)$$

$$\begin{aligned} H_T(2) &= \int_{-\infty}^{\infty} \frac{d\boldsymbol{k}}{(2\pi)^2} w c_b^\dagger(\boldsymbol{k}+\boldsymbol{g}_2) c_t(\boldsymbol{k}) + \mathrm{h.c.} \\ &= \int_{\mathrm{MBZ}} \frac{d\boldsymbol{k}}{(2\pi)^2} w \left[ c_b^\dagger(\boldsymbol{k}) c_t(\boldsymbol{k}-\boldsymbol{g}_2) + c_b^\dagger(\boldsymbol{k}-\boldsymbol{g}_3) c_t(\boldsymbol{k}+\boldsymbol{g}_1) \right] + \mathrm{h.c.} + (\text{other } \boldsymbol{k}), \end{aligned} \qquad (30)$$

$$\begin{aligned} H_T(3) &= \int_{-\infty}^{\infty} \frac{d\boldsymbol{k}}{(2\pi)^2} w c_b^\dagger(\boldsymbol{k}-\boldsymbol{g}_3) c_t(\boldsymbol{k}) + \mathrm{h.c.} \\ &= \int_{\mathrm{MBZ}} \frac{d\boldsymbol{k}}{(2\pi)^2} w \left[ c_b^\dagger(\boldsymbol{k}-\boldsymbol{g}_3) c_t(\boldsymbol{k}) + c_b^\dagger(\boldsymbol{k}+\boldsymbol{g}_1) c_t(\boldsymbol{k}-\boldsymbol{g}_2) \right] + \mathrm{h.c.} + (\text{other } \boldsymbol{k}). \end{aligned} \qquad (31)$$

Therefore, by keeping the states specified by $\mathbf{\Psi}_b(\mathbf{k})$ and $\mathbf{\Psi}_t(\mathbf{k})$, we obtain from $H_m^+$ and $H_\Delta^+$

$$H_\Delta^b = \int_{\text{MBZ}} \frac{d\mathbf{k}}{(2\pi)^2} \mathbf{\Psi}_b^\dagger(\mathbf{k}) \begin{pmatrix} 0 & Ve^{-i\psi} & Ve^{+i\psi} \\ Ve^{+i\psi} & 0 & Ve^{-i\psi} \\ Ve^{-i\psi} & Ve^{+i\psi} & 0 \end{pmatrix} \mathbf{\Psi}_b(\mathbf{k}) \equiv \int_{\text{MBZ}} \frac{d\mathbf{k}}{(2\pi)^2} \mathbf{\Psi}_b^\dagger(\mathbf{k}) h_\Delta^b(\mathbf{k}) \mathbf{\Psi}_b(\mathbf{k}), \tag{32}$$

$$H_\Delta^t = \int_{\text{MBZ}} \frac{d\mathbf{k}}{(2\pi)^2} \mathbf{\Psi}_t^\dagger(\mathbf{k}) \begin{pmatrix} 0 & Ve^{+i\psi} & Ve^{-i\psi} \\ Ve^{-i\psi} & 0 & Ve^{+i\psi} \\ Ve^{+i\psi} & Ve^{-i\psi} & 0 \end{pmatrix} \mathbf{\Psi}_t(\mathbf{k}) \equiv \int_{\text{MBZ}} \frac{d\mathbf{k}}{(2\pi)^2} \mathbf{\Psi}_t^\dagger(\mathbf{k}) h_\Delta^t(\mathbf{k}) \mathbf{\Psi}_t(\mathbf{k}), \tag{33}$$

$$H_T = \int_{\text{MBZ}} \frac{d\mathbf{k}}{(2\pi)^2} \mathbf{\Psi}_b^\dagger(\mathbf{k}) \begin{pmatrix} w & 0 & w \\ 0 & w & w \\ w & w & 0 \end{pmatrix} \mathbf{\Psi}_t(\mathbf{k}) + \text{h.c.} \equiv \int_{\text{MBZ}} \frac{d\mathbf{k}}{(2\pi)^2} \mathbf{\Psi}_b^\dagger(\mathbf{k}) h_{bt}(\mathbf{k}) \mathbf{\Psi}_t(\mathbf{k}) + \mathbf{\Psi}_t^\dagger(\mathbf{k}) h_{tb}(\mathbf{k}) \mathbf{\Psi}_b(\mathbf{k}). \tag{34}$$

Thus, we have rewritten Eq. (10) and obtain the moiré wave-number representation

$$H_{\tau=+} = \int_{\text{MBZ}} \frac{d\mathbf{k}}{(2\pi)^2} \left( \mathbf{\Psi}_b^\dagger(\mathbf{k}), \mathbf{\Psi}_t^\dagger(\mathbf{k}) \right) H_{\tau=+}(\mathbf{k}) \begin{pmatrix} \mathbf{\Psi}_b(\mathbf{k}) \\ \mathbf{\Psi}_t(\mathbf{k}) \end{pmatrix}, \tag{35}$$

$$H_{\tau=+}(\mathbf{k}) \equiv \begin{pmatrix} h_m^b(\mathbf{k}) + h_\Delta^b(\mathbf{k}) & h_{bt}(\mathbf{k}) \\ h_{tb}(\mathbf{k}) & h_m^t(\mathbf{k}) + h_\Delta^t(\mathbf{k}) \end{pmatrix}. \tag{36}$$

We can also obtain that for $\tau = -$ and the BdG Hamiltonian in the same way.

### Details of the numerical calculations and the sign reversal of $\sigma_{\text{inj}}^{y;yy}$

Here, we discuss the details of the numerical calculatioins. In particular, we explain how the sign reversal of $\sigma_{\text{inj}}^{y;yy}$ in Fig. 4(a) occurs in the nematic state.

For numerical calculations, delta functions in JDOS and $\sigma_{\text{inj}}^{y;yy}$ are approximated by the Lorentzian $\mathcal{L}(x) = \frac{1}{\pi} \frac{\Gamma}{x^2 + \Gamma^2}$. Here, following Ref. [9], we adapted the same value of $\Gamma$ as the scattering rate, which is introduced by replacing the adiabaticity parameter $\eta$ of $\sigma_{\text{inj}}^{y;yy}$ as discussed in the main text. We determined the value of the scattering rate $\Gamma$ in the following way. According to Ref. [3], the mean-free path and coherence length are $l = 229$nm and $\xi_0 = 57$nm, where $l \gg \xi_0$ indicates that the system is in the clean limit. By assuming the relations $\xi_0 = \hbar v_F / \pi \Delta_0$, $l = v_F / \Gamma_N$, and with a typical value of the Fermi velocity $v_F$, we obtain the normal-state scattering rate $\Gamma_N \sim 50.5 \mu$eV, which amounts to $\sim \Delta_0/2$. The scattering rate in the superconducting state would be smaller than $\Gamma_N$ due to the gap opening, and thus we phenomenologically assume a smaller value $\Gamma = \Delta_0/50 < \Gamma_N$. We have confirmed that the obtained response is suppressed according to $\sigma_{\text{inj}}^{y;yy} \sim 1/\Gamma$ by increasing $\Gamma$ and keep the qualitative temperature dependence intact, as long as $\Gamma$ is not too large.

As discussed in the main text, $\sigma_{\text{inj}}^{y;yy}$ of the nematic state shows a sign reversal near $\Omega \sim 110 \mu$eV. This behavior can be understood from Eq. (8) and the distributions of the effective mass shown in Fig. 5(b). Figure 5(a) shows the $\mathbf{k}$ dependence of the excitation energy between the quasiparticle bands $A, B$, i.e., $E_A(\mathbf{k} + \mathbf{Q}/2) - E_B(\mathbf{k} - \mathbf{Q}/2)$, in the presence of a finite momentum transfer $\mathbf{Q}$. It is evident that the main contributions to the resonance originate from regions near the Fermi surfaces of valley $+$ in the normal state, which consist of a large open Fermi surface and small Fermi pockets near the moiré K point as indicated by black lines in Fig. 5(b). Note that the acoustogalvanic conductivity $\sigma_{\text{inj}}^{y;yy}$ can be understood as the summation of the contributions by these large and small Fermi surfaces. According to Eq. (8), the sign of the contribution from each Fermi surface is determined by the effective mass. While the large Fermi surface does not have a definite sign of the effective mass due to the presence of sign reversals, it turns out that the region of $m^{yy}(\mathbf{k}) < 0$ mainly contributes for the frequency under consideration. Similarly, the small Fermi pockets can be regarded to have positive effective mass. Thus, their contributions to $\sigma_{\text{inj}}^{y;yy}$ are of different signs. This can be confirmed from Fig. 5(c), where the contributions from large and small Fermi surfaces, indicated by blue and red lines, respectively, have opposite signs. The dominant contribution to the acoustogalvanic conductivity is coming from the small and large Fermi surfaces for small and large frequencies, thereby explaining the sign reversal of $\sigma_{\text{inj}}^{y;yy}$.

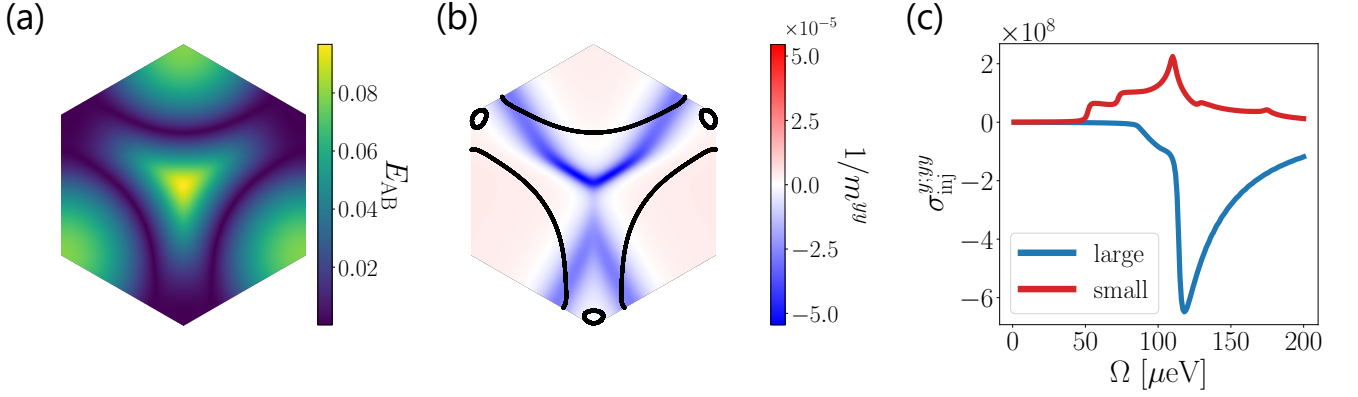

FIG. 5. Color map of (a) $E_{AB} = E_A(\mathbf{k}+\mathbf{Q}/2) - E_B(\mathbf{k}-\mathbf{Q}/2)$ and (b) the inverse effective mass $1/m^{yy}(\mathbf{k})$ of the energy band of the + valley near the Fermi energy. In panel (b), the black lines represent the Fermi surfaces of the valley +. (c) Contribution of each Fermi surfaces to $\sigma^{y;yy}_{\text{inj}}$. The blue and red lines correspond to large and small Fermi surfaces.

### Formulation of SAGE

Here, we discuss the derivation of the formulas of acoustogalvanic conductivity. In the presence of pseudo-gauge field, the normal-state Hamiltonian of the valley $\tau$ is written as

$$\hat{H}_\tau[\mathbf{A}_s] = \hat{H}_{\tau,l}[0] - \int \frac{d^2q}{(2\pi)^2} \hat{V}^{(\beta,\mathbf{q})}_\tau A^\beta_s(\mathbf{q},t) + \frac{1}{2}\int \frac{d^2q}{(2\pi)^2}\int \frac{d^2q'}{(2\pi)^2} \hat{V}^{(\beta,\mathbf{q})(\gamma,\mathbf{q}')}_\tau A^\beta_s(\mathbf{q},t)A^\gamma_s(\mathbf{q}',t) + O(A_s^3), \quad (37)$$

by using the Fourier coefficients of the pseudo-gauge field,

$$\mathbf{A}_s(\mathbf{r},t) = \int \frac{d^2q}{(2\pi)^2} e^{i\mathbf{q}\cdot\mathbf{r}} \mathbf{A}_s(\mathbf{q},t) = \int \frac{d^2q}{(2\pi)^2}\int \frac{d\omega}{2\pi} e^{i\mathbf{q}\cdot\mathbf{r}-i\omega t} \mathbf{A}_s(\mathbf{q},\omega). \quad (38)$$

Here, the current operators for the pseudo-gauge field is given by $\hat{V}^{(\beta,\mathbf{q})}_\tau = \text{diag}(\hat{V}^{(\beta,\mathbf{q})}_{\tau,b}, \hat{V}^{(\beta,\mathbf{q})}_{\tau,t})$ and $\hat{V}^{(\beta,\mathbf{q})(\gamma,\mathbf{q}')}_\tau = \text{diag}(\hat{V}^{(\beta,\mathbf{q})(\gamma,\mathbf{q}')}_{\tau,b}, \hat{V}^{(\beta,\mathbf{q})(\gamma,\mathbf{q}')}_{\tau,t})$, with

$$\hat{V}^{(\beta,\mathbf{q})}_{\tau,l} = \left\{ -\frac{\tau\hbar^2}{2m^*}(\hat{\mathbf{k}} - \boldsymbol{\kappa}_{\tau,l}),\ e^{i\mathbf{q}\cdot\hat{\mathbf{r}}} \right\}, \qquad \hat{V}^{(\beta,\mathbf{q})(\gamma,\mathbf{q}')}_{\tau,l} = -\frac{\hbar^2\tau^2}{m^*} e^{i(\mathbf{q}+\mathbf{q}')\cdot\hat{\mathbf{r}}} \delta_{\beta\gamma}. \quad (39)$$

Similar expansion is obtained for the BdG Hamiltonian:

$$\hat{H}_{\text{BdG}}[\mathbf{A}_s] = \hat{H}_{\text{BdG}}[0] - \int \frac{d^2q}{(2\pi)^2} \hat{V}^{(\beta,\mathbf{q})} A^\beta_s(\mathbf{q},t) + \frac{1}{2}\int \frac{d^2q}{(2\pi)^2}\int \frac{d^2q'}{(2\pi)^2} \hat{V}^{(\beta,\mathbf{q})(\gamma,\mathbf{q}')} A^\beta_s(\mathbf{q},t) A^\gamma_s(\mathbf{q}',t) + O(A_s^3), \quad (40)$$

with $\hat{V}^{(\beta,\mathbf{q})} = \text{diag}(\hat{V}^{(\beta,\mathbf{q})}_+, -[\hat{V}^{(\beta,\mathbf{q})}_-]^T)$, and $\hat{V}^{(\beta,\mathbf{q})(\gamma,\mathbf{q}')} = \text{diag}(\hat{V}^{(\beta,\mathbf{q})(\gamma,\mathbf{q}')}_+, \hat{V}^{(\beta,\mathbf{q})(\gamma,\mathbf{q}')}_-)$. For the latter use, we define

$$\hat{V}^\beta \equiv \hat{V}^\beta(\mathbf{q}=0), \qquad \hat{V}^{(\beta,\mathbf{q})} = \frac{1}{2}\left\{ V^\beta, e^{i\mathbf{q}\cdot\hat{\mathbf{r}}} \right\}. \quad (41)$$

The acoustogalvanic conductivity can be derived by solving the von-Neumann equation for the density matrix with the perturbation $\mathbf{A}_s(\mathbf{r},t)$, in the same way as the nonlinear optical responses. For this purpose, it is convenient to rewrite Eq. (40) in the following form by generalizing the Einstein summation rule:

$$\hat{H}_{\text{BdG}}[\mathbf{A}_s] = \hat{H}_{\text{BdG}} - \hat{V}^{\bar{\beta}} A^{\bar{\beta}}_s(t) + \frac{1}{2}\hat{V}^{\bar{\beta}\bar{\gamma}} A^{\bar{\beta}}_s(t) A^{\bar{\gamma}}_s(t) + O(A_s^3). \quad (42)$$

Here, $\bar{\beta} = (\beta, \mathbf{q})$ and

$$\hat{V}^{\bar{\beta}} A^{\bar{\beta}}_s(t) \equiv \sum_\beta \int \frac{d^2q}{(2\pi)^2} \hat{V}^{(\beta,\mathbf{q})} A^\beta_s(\mathbf{q},t), \quad (43)$$

for example. By using this notation, we can immediately obtain the acoustogalvanic conductivity based on the results of the optical conductivity [9]. This is achieved by replacing the gauge-field derivative $\partial_{A^\beta}$ and $\partial_{A^\gamma}$ with the pseudo-gauge-field derivative $\partial_{A_s^{\bar\beta}}$ and $\partial_{A_s^{\bar\gamma}}$. Thus, the acoustogalvanic conductivity corresponding to the injection-current optical response is given by

$$\sigma_{\boldsymbol{Q},-\boldsymbol{Q}}^{\alpha;\beta\gamma}(0;\Omega,-\Omega) = \sigma^{\alpha;\bar\beta,\bar\gamma}(0;\Omega,-\Omega) = \frac{1}{\Omega^2}\frac{-\pi}{2\eta V}\sum_{ac}\delta\mathcal{J}_{ac}^\alpha V_{ac}^{\bar\beta}V_{ca}^{\bar\gamma}f_{ac}\delta(\Omega-E_{ac}), \quad \bar\beta=(\beta,\boldsymbol{Q}), \quad \bar\gamma=(\gamma,-\boldsymbol{Q}), \quad (44)$$

with the system area $V$. The removal of the overall coefficient $1/2$ from the expression in Ref. [9] is due to the different conventions of the BdG Hamiltonian: That in Ref. [9] has a twice larger matrix size than that in this paper. In this expression, the indices $a$ and $c$ independently run over all the eigenstates of $\hat{H}_{\rm BdG}$ including the wave number. The quantities in the summand are defined by $V_{ac}^{\bar\beta} \equiv \left\langle \psi_a \middle| \hat{V}^{\bar\beta} \middle| \psi_c \right\rangle$ with $\hat{H}_{\rm BdG}\ket{\psi_a} = E_a\ket{\psi_a}$, as well as $\delta\mathcal{J}_{ac}^\alpha = \mathcal{J}_{aa}^\alpha - \mathcal{J}_{cc}^\alpha$ with $\mathcal{J}_{aa}^\alpha = \left\langle \psi_a \middle| \hat{\mathcal{J}}^\alpha \middle| \psi_a \right\rangle$. We defined the spatially-uniform charge current operator $\hat{\mathcal{J}}^\alpha$ by the gauge-field derivative of the BdG Hamiltonian [9]. In particular, we can write $\hat{\mathcal{J}}^\alpha = \hat{V}^\beta\tau_z$, with the Pauli matrix $\tau_z$ in the Nambu space.

Let us redefine $a$ and $c$ to represent the band indices of the BdG Hamiltonian, and $a \to (\boldsymbol{k}_a, a)$ and $c \to (\boldsymbol{k}_c, c)$. Then, the matrix element $V_{ac}^{\bar\beta}$ reduces to

$$V_{ac}^{\bar\beta} = \left\langle \psi_a(\boldsymbol{k}_a) \middle| \hat{V}^{(\beta,\boldsymbol{Q})} \middle| \psi_c(\boldsymbol{k}_c) \right\rangle = \delta_{\boldsymbol{k}_a,\boldsymbol{k}_c+\boldsymbol{Q}}V_{ac}(\boldsymbol{k}_a,\boldsymbol{k}_c), \quad (45)$$

with

$$V_{ac}^\beta(\boldsymbol{k}_a,\boldsymbol{k}_c) \equiv \left\langle u_a(\boldsymbol{k}_a) \middle| \frac{\hat{V}^\beta(\boldsymbol{k}_a)+\hat{V}^\beta(\boldsymbol{k}_c)}{2} \middle| u_c(\boldsymbol{k}_c) \right\rangle. \quad (46)$$

We defined $\hat{V}^\beta(\boldsymbol{k}) = e^{-i\boldsymbol{k}\cdot\hat{\boldsymbol{r}}}\hat{V}^\beta e^{i\boldsymbol{k}\cdot\hat{\boldsymbol{r}}}$ and denoted the Bloch state for the BdG Hamiltonian by $\ket{\psi_a(\boldsymbol{k})} = e^{i\boldsymbol{k}\cdot\hat{\boldsymbol{r}}}\ket{u_a(\boldsymbol{k})}$ with its periodic part $\ket{u_a(\boldsymbol{k})}$. By defining

$$\mathcal{J}_{aa}^\alpha(\boldsymbol{k}_a) \equiv \left\langle u_a(\boldsymbol{k}_a) \middle| \hat{\mathcal{J}}^\alpha(\boldsymbol{k}_a) \middle| u_a(\boldsymbol{k}_a) \right\rangle, \quad \hat{\mathcal{J}}^\alpha(\boldsymbol{k}_a) = e^{-i\boldsymbol{k}_a\cdot\hat{\boldsymbol{r}}}\mathcal{J}^\alpha e^{i\boldsymbol{k}_a\cdot\hat{\boldsymbol{r}}}, \quad (47)$$

and introducing $\boldsymbol{k} = \boldsymbol{k}_c + \boldsymbol{Q}/2$, we obtain

$$\begin{aligned}\sigma_{\rm inj}^{\alpha;\beta\gamma} &= \frac{\sigma^{\alpha;\bar\beta,\bar\gamma}(0;\Omega,-\Omega)+(\beta\leftrightarrow\gamma)}{2} \\ &= \frac{1}{2}\frac{-\pi}{2\eta V}\sum_{\boldsymbol{k},\boldsymbol{k}_a,\boldsymbol{k}_c,a,c}\delta_{\boldsymbol{k}_a,\boldsymbol{k}+\frac{\boldsymbol{Q}}{2}}\delta_{\boldsymbol{k}_c,\boldsymbol{k}-\frac{\boldsymbol{Q}}{2}}(\mathcal{J}_{aa}^\alpha(\boldsymbol{k}_a)-\mathcal{J}_{cc}^\alpha(\boldsymbol{k}_c))\frac{V_{ac}^\beta(\boldsymbol{k}_a,\boldsymbol{k}_c)V_{ca}^\gamma(\boldsymbol{k}_c,\boldsymbol{k}_a)}{(E_a(\boldsymbol{k}_a)-E_c(\boldsymbol{k}_c))^2} \\ &\quad \cdot \Big(f(E_a(\boldsymbol{k}_a))-f(E_c(\boldsymbol{k}_c))\Big)\delta\big(\Omega-E_a(\boldsymbol{k}_a)+E_c(\boldsymbol{k}_c)\big)+(\beta\leftrightarrow\gamma).\end{aligned} \quad (48)$$

By replacing $(1/V)\sum_{\boldsymbol{k}}$ with $\int\frac{d^2k}{(2\pi)^2}$, we obtain the formula in the main text.


\fi

\end{document}